%% file: pne_apss_final.tex
\documentclass{aastex}          
\usepackage{spr-astr-addons}    

\usepackage{hyperref}
\usepackage{graphicx}
\usepackage{appendix}
\usepackage{psfrag}
\usepackage{amsmath,amssymb}
\usepackage{threeparttable}
\usepackage{float}
\usepackage{subfigure}
\usepackage{afterpage}
\usepackage{longtable,lscape}
\usepackage{multicol}
\usepackage{rotating}
\usepackage{adjustbox}
\usepackage{hyperref}
\usepackage{pdfpages}
\usepackage{epstopdf}
\usepackage{tabularx}
\def\aaps{A\&AS}
\def\apj{ApJ}
\def\apjl{ApJL}
\def\mnras{MNRAS}

\def\pasp{PASP}
\def\aj{AJ}

\def\aap{A\&A}
\def\apjs{ApJS}

\def\mnras{MNRAS}

\begin{document}
%
\title{A catalogue of 108 extended planetary nebulae observed by \textit{GALEX}}

\shorttitle{GALEX observations of planetary nebulae}
\shortauthors{Pradhan et al.}


\author{Ananta C. Pradhan\altaffilmark{1}} \and \author{Swayamtrupta Panda\altaffilmark{2,3}}  \and \author{Mudumba Parthasarathy\altaffilmark{4,5}} \and \author{Jayant Murthy\altaffilmark{4}} \and \author{Devendra K. Ojha\altaffilmark{6}}
\email{acp.phy@gmail.com} 

\altaffiltext{1}{Department of Physics \& Astronomy, National Institute of Technology, Rourkela, Odisha - 769008, India \\\href{acp.phy@gmail.com}{(Email: acp.phy@gmail.com)}}
\altaffiltext{2}{Center for Theoretical Physics, Al. Lotnik\'ow 32/46, 02-668 Warsaw, Poland}
\altaffiltext{3}{Nicolaus Copernicus Astronomical Center, ul. Bartycka 18, 00-716 Warsaw, Poland}
\altaffiltext{4}{Indian Institute of Astrophysics, II Block, Koramangala, Bangalore, 560 034, India}
\altaffiltext{5}{National Astronomical Observatory of Japan (NAOJ), 2-21-1 Osawa, Mitaka, Tokyo 181-8588, Japan}
\altaffiltext{6}{Tata Institute of Fundamental Research, Homi Bhabha Road, Colaba, Mumbai, 400005, India}


\begin{abstract}
We present the ultraviolet (UV) imaging observation of planetary nebulae (PNe) using archival data of \textit{Galaxy Evolution Explorer (GALEX)}. We found 358 PNe detected by \textit{GALEX} in near-UV (NUV). We have compiled a catalogue of 108 extended PNe with sizes greater than 8\arcsec and provided the angular diameters for all the 108 extended PNe in NUV and 28 in FUV from the \textit{GALEX} images considering 3$\sigma$ surface brightness level above the background. Of the 108 PNe, 74 are elliptical, 24 are circular and 10 are bipolar in NUV with most being larger in the UV than in the radio, H$\alpha$ or optical. We derived luminosities for 33 PNe in FUV (L$_{\text{FUV}}$) and 89 PNe in NUV (L$_{\text{NUV}}$) and found that most of the sources are very bright in UV. The FUV emission of the \textit{GALEX} band includes contribution from prominent emission lines N IV] (1487 \AA), C IV (1550 \AA), and O III] (1661 \AA) whereas the NUV emission includes C III] (1907 \AA) and C II (2325 \AA) for PNe of all excitation classes. The other emission lines seen in low excitation PNe are O IV] (1403 \AA) and N III (1892 \AA) in FUV, and O II (2470 \AA) and Mg II (2830 \AA) in NUV. Similarly the emission lines O V (1371 \AA) and He II (1666 \AA) strongly contribute in FUV for high and medium excitation PNe but not for low excitation PNe. A mixture of other emission lines seen in all excitation PNe. We have also provided images of 34 PNe in NUV and 9 PNe in FUV.
 
\end{abstract}



\section{Introduction}

The low and intermediate mass stars (1 - 8 M$_\odot$) during the advanced stage of evolution go through the planetary nebula (PN) phase. A PN consists of a hot, luminous central star and an expanding glowing shell of gas and dust. More than 3000 true and probable planetary nebulae (PNe) have been detected in the Milky Way which are catalogued by \citet{Acker92} and \citet{Parker06}. The central star of most of the PNe (CSPNe) is a very hot object (T $> 30$ kK) which is bright in ultraviolet (UV) and hence, UV observations explore the properties of these objects. The intense UV light irradiates the material at the outer shell of the stars and ionizes the atoms in the gas.

\par 
The appearance of the spectrum of a PN depends upon temperature, luminosity, and chemical composition. \citet{Mendez91} suggested that the majority of CSPNe can be classified in two distinct categories: those for which stellar H features can be identified in their spectra (hydrogen-rich) and those for which they cannot (hydrogen-poor). The hydrogen-rich CSPNe become DA-type white dwarfs while the hydrogen-poor become DO-type white dwarfs at the end of their evolutionary life \citep{Werner06}. The determination of spectral types of CSPNe should help significantly to improve our knowledge of their general evolutionary scheme, making it possible to consider CSPNe as physical objects with individual parameters and peculiarities and not just as sources of ionizing radiation \citep{Weid11}. 
\par 
The UV spectral region of PNe contains important nebular emission lines  due to carbon, nitrogen,  silicon, neon, argon, etc., and P-Cygni type stellar wind profiles of C IV, N V, etc., from the hot CSPNe. Forbidden lines from ions such as the three and four times ionized neon and argon enable one to probe the high temperature regions of the PNe. UV spectra of large number of PNe were studied with the \textit{International Ultraviolet Explorer (IUE)} satellite covering the wavelength region 1150 \AA\ to 3200 \AA\ \citep[][references therein]{Boggess80, Koppen87, Heap97, Pauldrach04, Herald11, Keller14} and with the \textit{Far-Ultraviolet Spectroscopic Explorer (FUSE)} covering the wavelength region 905 \AA\ to 1187 \AA\ \citep{Heap87, Guerrero13, Herald11, Keller14}. The important spectral lines in the UV spectra of PNe are given in the above mentioned papers. \citet{Boggess80} have also given absolute fluxes of several UV nebular emission lines for several PNe with excitation class ranging from 1 to 10 with 10 being the highest. Recently, \citet{Rao18a,Rao18b} have studied UV structure of PNe NGC 6302 and NGC 40 using observation of \textit{Ultraviolet Imaging Telescope (UVIT)} onboard \textit{ASTROSAT}.

Since CSPNe are hot post-AGB stars, they are easily detected in the UV than in the optical images. Also, the morphology of PNe in UV can be different from that in optical as UV traces the hot gases in the inner regions of the nebulae. It is important to compare the UV images with their optical counterparts to understand the wavelength dependent morphology of PNe. Many PNe show bi-polar structures, collimated outflows, jets and knots. The morphological classification of PNe depends upon these structural appearances in the imaging surveys. In recent years, several attempts have therefore been made to classify the PNe samples observed by various narrow and broad band imaging surveys \citep{Bal87, Aaquist96, Man96, Man00,Parker06, Kwok10, Stan16, Weid16}. 
\par 

 The \textit{Galaxy Evolution Explorer (GALEX)}\footnote{\href{https://archive.stsci.edu/missions-and-data/galex-1/}{https://archive.stsci.edu/missions-and-data/galex-1/}} UV sky surveys in far-UV (FUV) and near-UV (NUV) have not only observed a good census of hot source candidates of the Milky Way \citep{Bianchi11, Pradhan14} but also observed a significant sample of PNe. The \textit{GALEX} images of selected PNe contain the nebular UV emission (continuum  $+$ emission lines) and the UV continuum from CSPNe \citep{Bianchi12b}. In some cases the contribution from the He II 1640 \AA\ line is also present. The FUV images do not include the N V line. At 5\arcsec\;resolution of \textit{GALEX}, the FUV and NUV images of selected PNe are useful for studies of large PNe where the deep sensitivity combined with low sky background in FUV may reveal faint structures, shock fronts, collimated outflows and morphology. \citet{Bianchi12a} have presented \textit{GALEX} color-composite images for a sample of 13 PNe.

The angular dimension is one of the most important observational quantities used to characterize PN \citep{van00}. They are required for the study of crucial PN parameters such as linear dimensions, lifetimes, distances and masses. Different observational techniques and methods for measuring sizes give varied results for the PNe. Several methods such as direct measurements at 10\% level of the peak surface brightness, Gaussian deconvolution, second-moment deconvolution, etc., have been used to measure the angular dimensions of observed PNe \citep{Tylenda03}. \citet{van00} has made a thorough model analysis of these methods for measuring partially resolved PNe.

\section{Observations}

The \textit{GALEX} imaging surveys have scanned the sky in FUV (1344 - 1786 \AA, $\lambda_\mathrm{eff}$ = 1538.6 \AA) and NUV (1771 - 2831 \AA, $\lambda_\mathrm{eff}$ = 2315.7 \AA) bands with a spatial resolution of 4.2\arcsec\ and 5.3\arcsec, respectively. Each \textit{GALEX} image has a 1.25\degr\ field of view and the size of each pixel in an image is about 1.5\arcsec. The details about \textit{GALEX} is described by \citet{Morrissey07}. The final release of \textit{GALEX} data is available in Mikulski Archive for Space Telescope (MAST)\footnote{\href{http://archive.stsci.edu/}{http://archive.stsci.edu/}}. This has covered about 75\% of the sky including the bright Galactic bulge and the Magellanic clouds in a decade of \textit{GALEX} operation. Due to its high resolution and deep sensitivity, \textit{GALEX} data are well suited to study the large extended objects like PNe. 

\begin{figure*}
\begin{center}
\includegraphics[width=15.5cm]{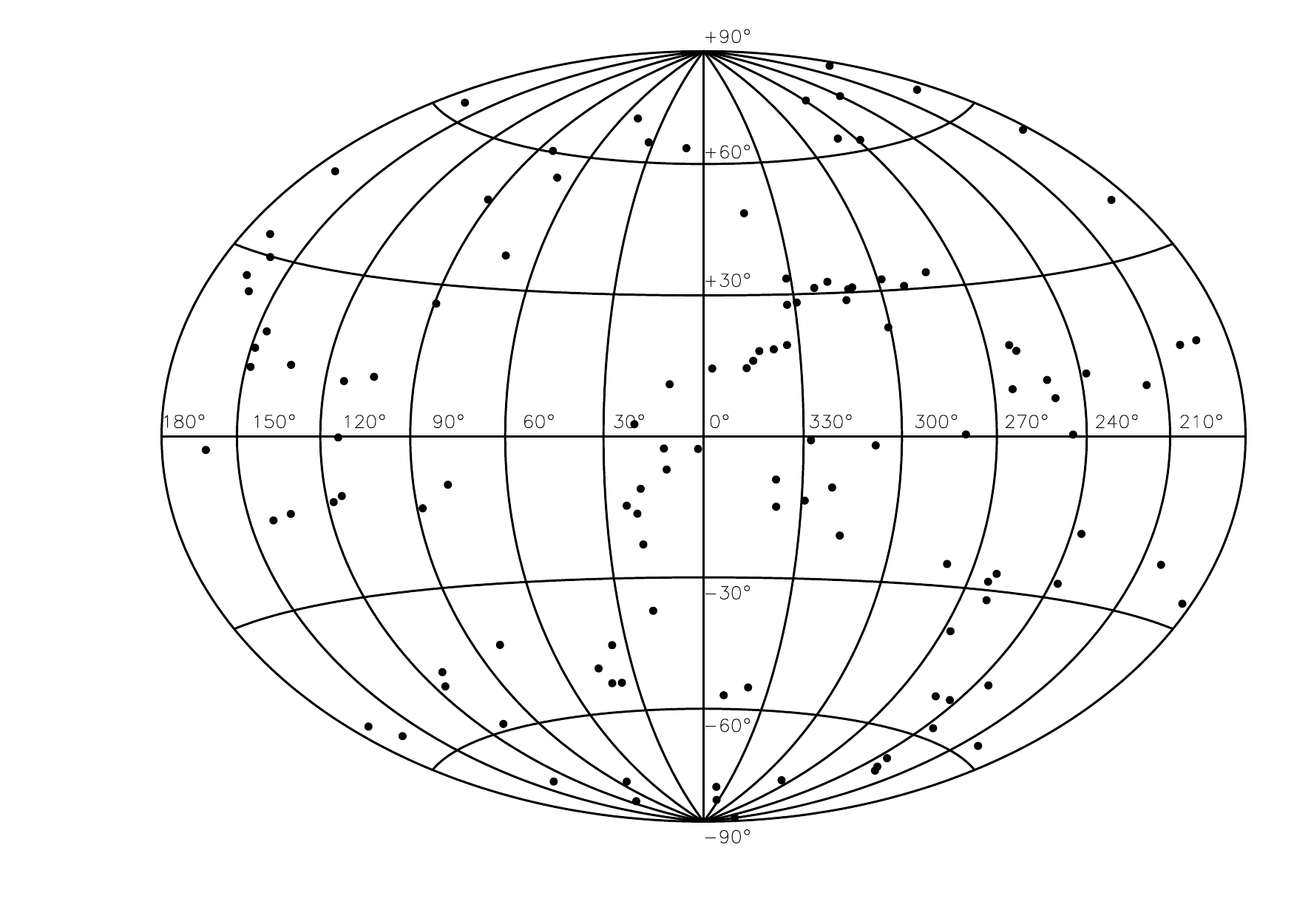}
\caption{Distribution of the 108 extended PNe detected in NUV band of \textit{GALEX} is shown in an Aitoff projection in Galactic coordinates.}
\label{fig1}
\end{center}
\end{figure*}

 We downloaded the Strasbourg-ESO catalogue of Galactic PNe by \citet{Acker92} and the MASH catalogue of PNe by \citet{Parker06}. We used \textit{GALEX} CASjobs to search for their \textit{GALEX} detections in the \textit{GALEX} database using a search radius of 5\arcsec. We found 358 PNe which have at least a \textit{GALEX} NUV detection. Most of the PNe that we found are compact for the \textit{GALEX} resolution. So, we have chosen only the extended PNe which have a diameter larger than 8\arcsec. Then we were left with 108 PNe in which 92 are from \citet{Acker92} catalogue while 16 are from the MASH catalogue. Most of these objects are from All Sky Imaging Survey (AIS), except a few which are from Medium Sky Survey (MIS), Nearby Galaxy Survey (NGS), Nearby Imaging Survey (NIS) and GII survey of \textit{GALEX}. For some of the PNe, the CSPNe is not hot enough to be detected in UV. However, 28 of the extended PNe in our list have both the FUV and NUV detections. We have shown all the 108 PNe in an Aitoff projection in Galactic coordinates (Figure \ref{fig1}). The details of the 108 PNe have been consolidated in Table \ref{tab1}. The description of columns in Table \ref{tab1} are:
\begin{itemize}
\item \textbf{Col. 1} - the generic PN G name taken from the Strasbourg-ESO Catalogue \citep{Acker92} and the MASH catalogue \citep{Parker06} of Galactic PNe;
\item \textbf{Col. 2} - the common name as tabulated in MASH catalogue \citep{Parker06} and \citet{Weid11};
\item \textbf{Cols. 3-4} - the equatorial coordinates (J2000) of the PNe;
\item \textbf{Col. 5} - the extinction values from \citet{Schlegel98};
\item \textbf{Cols. 6-7} - the corresponding nebular NUV and FUV luminosities of the PNe (extinction-corrected);
\item \textbf{Col. 8} - the spectral classification of the CSPNe are taken from \citet{Weid11};
\item \textbf{Cols. 9-10} - the estimated NUV and FUV angular sizes in arcsecond;
\item \textbf{Col. 11} - the optical (V-band) size in arcsecond obtained from \citet{Acker92};
\item \textbf{Col. 12} - the H$\alpha$ size in arcsecond obtained from \citet{Frew16}; and
\item \textbf{Col. 13} - the morphology of the PNe. This is a proper classification.
\end{itemize}

\section{Results and Discussions}

 \subsection{Size of the PNe in UV}
 
  A number of observational surveys have measured the dimensions and structures of PNe at different wavelengths \citep{Acker92, Tylenda03, Ruffle04, Frew16}. However, the intrinsic morphology of PNe is not easily derived from the observed images as it depends on factors such as the sensitivity of the observing instruments, ionization structures, and the projection effect \citep{Kwok10}. PNe are mainly studied in optical wavelengths (for e.g., \citet{Frew16} and the reference therein), however, multi-wavelength observations are required to see how the various structural components of PNe are manifesting at different wavelengths.
  
We have determined the angular sizes for 108 PNe in NUV and for 28 PNe in FUV. Initially, we used the direct measurements at the 10\% level of the peak surface brightness but the \textit{GALEX} images are very deep and the nebular emission extends beyond 10\% contour. Hence, we measured the actual dimension of the nebula considering a 3$\sigma$ emission level above the background. The NUV and FUV sizes (measured in arcseconds) are provided in columns 9 and 10 of Table \ref{tab1}, respectively. The values denoted are mostly in the format $a \times b$, where $a,b$ are the semi-major and semi-minor axes respectively. For some of the PNe candidates, the single value of angular sizes refers to the round type PNe.
\par
The effect of projection must be also taken into account when extrapolating the two-dimensional information for the PN images to interpret the three-dimensional morphology. A supposed prolate ellipsoidal PN could be mistaken for a round, symmetric type when viewed from an angle that is different (i.e., edge-on for a face-on and vice-versa). The same may occur for a bipolar (B) type viewed face-on which may cast a round (R) or elliptical (E) structure to the observer. In order to find out the degree to which E-type and B-type PNe could be confused with R-type PNe, the probability of mistaking one of these PNe as R-type has been taken into consideration. For instance, the probability of perceiving an E-type as a R-type depends on the axial ratio, the inclination angle, the size, and the spatial resolution. According to the average size of R types and our spatial resolution, we have classified R types as having axial ratios $a/b \le 1.06$ as per the scheme in \citet{Man04}.
\par
Taking the method proposed by \citet{Man04} a step further, we try to classify our nebulae based on their NUV sizes either as round or elliptical. This method uses only the values obtained for the angular extensions as semi-major and semi-minor axes, and thus ignores the visual detection leading to categorising some PNe as bipolar. Here, the morphology of PNe is estimated from the axial ratio $a/b$. If $0.94 \le a/b \le 1.06$ then the PNe is said to be round and if $a/b > 1.06$ it is marked as an elliptical. We have therefore 24 round-type and 74 elliptical identified by this analysis. With our visual classifications, the other 10 candidates are classified as bipolar. These morphological classes are provided in the 13$^{\text{th}}$ column of Table \ref{tab1}.

\subsection{Interpreting wavelength vs. size of PNe}

The envelope surrounding the central star of PNe consists of ionized, atomic, molecular, and dust components. These components exist in a range of temperature from 100 K to 10$^{6}$ K. Hence the PNe radiate strongly throughout the electromagnetic spectrum from radio to X-ray and the dimensions of the PNe have been measured in all these wavelengths. The dimensions of the PNe solely depend upon the maximum extension of the respective recorded emission. Larger extent of the envelope is determined by the peak wavelength which is inversely related to the effective temperature. The variation of the angular size with respect to effective wavelength is a relative measure of the extent of hotness of the CSPNe.

\begin{figure*}
\centering
\begin{center}
\includegraphics[width=10cm, height=15.5cm, angle=270]{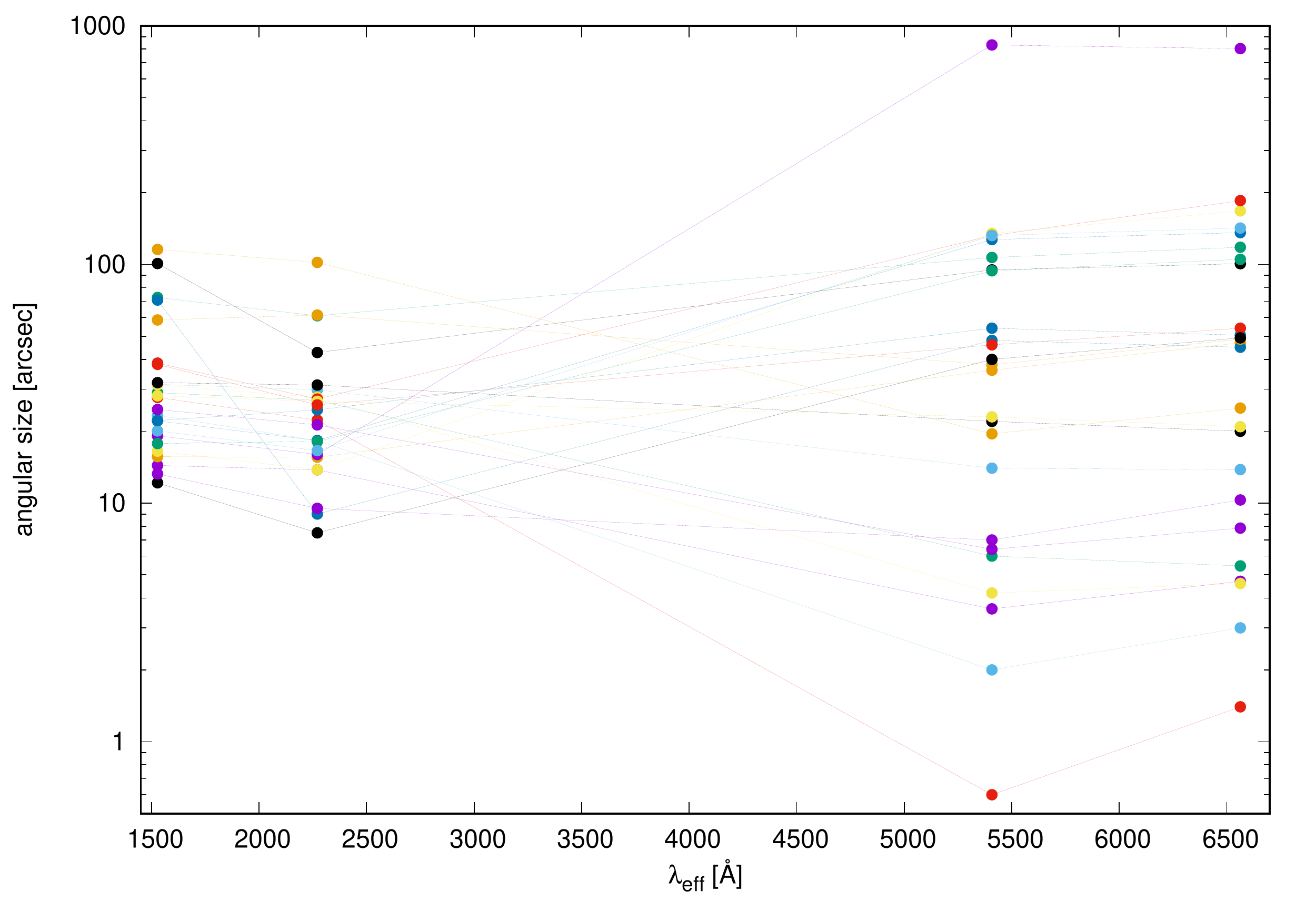}
\caption{Angular size distribution for 24 planetary nebulae across 4 distinct wavelengths regimes: FUV (1528 \AA), NUV (2271 \AA), V-Band (5408 \AA) and H$\alpha$ (6562.8 \AA). The angular sizes are taken from this paper (FUV and NUV), \citet{Acker92} (V-Band) and \citet{Frew16} (H$\alpha$).}
\label{fig2}
\end{center}
\end{figure*}

Out of the 108 PNe, the data for the angular sizes are available for 81 in optical (V band) \citep{Acker92}, 48 in H$\alpha$ \citep{Frew16} and 28 in FUV band of \textit{GALEX} (Table \ref{tab1}). We found 24 PNe in the list which have angular sizes data in four wavelength bands i.e., in FUV (1538.6 \AA), NUV (2315.7 \AA), V (5448 \AA) and H$\alpha$ (6562.8 \AA). 
We do not see any systematic trend in variation of size with wavelength (Figure \ref{fig2}). The PNe show a clear-cut decrease in sizes from FUV to NUV indicating that the PNe are brighter in FUV than NUV band. However, this decreasing trend is not followed further with increase in wavelengths for all the PNe.  Again there is a mild increase in sizes seen from V to H$\alpha$ band for almost all 24 PNe as the envelope of all PNe are rich in strong H$\alpha$ emission lines. The sizes in all four bands are approximately in a span of 10 to 100 arcsecs except a few deviations.

 \subsection{Intrinsic luminosities of PNe}
 
The integrated fluxes of the PNe in FUV and NUV bands of \textit{GALEX} are calculated using the methods described in \cite{Kwok08} and \cite{Zhang12}. Two individual diaphragms of same sizes have been considered  for the purpose:  one is used to measure the flux (FN) on the nebula covering the entire nebular extension in the image and the other diaphragm is used to measure the background flux. The background flux has been calculated at three different random blank locations of each image to minimize the errors and then the average of the three measurements is subtracted from FN to obtained the observed total integrated nebular flux of the PNe. This integrated flux includes the flux of the central star,  nebular emission and the contribution from the emission lines.

 The observed UV flux for each of the planetary nebulae includes the intrinsic flux and the effect of interstellar extinction. The visual extinction for the PNe is taken from \citet{Schlegel98} maps and then using the extinction law from \citet{Car89}, the extinction coefficients, A$_{\text{FUV}} = 8.16 \times \text{E(B-V)}$ and A$_{\text{NUV}} = 8.90 \times \text{E(B-V)}$ have been calculated. The intrinsic FUV and NUV luminosities (L$_{\text{FUV}}$ and L$_{\text{NUV}}$) for the PNe are calculated from the intrinsic fluxes (F$_{\text{FUV}}$ and F$_{\text{NUV}}$) obtaining the nebular distances from  \textit{GAIA} database\footnote{https://gea.esac.esa.int/archive/} \citep{Gaia18}, \cite{Frew16} and \citet{Acker92}. 
 We have determined L$_{\text{FUV}}$ for 33 PNe and L$_{\text{NUV}}$ for 89 PNe depending on the availability of distance for our sample (Table \ref{tab1}). The luminosities for the PNe are found to lie within the range of $8.02\times 10^{29}$ erg/s to $4.72\times 10^{35}$ erg/s in FUV and $2.71\times 10^{28}$ erg/s to $2.01\times 10^{36}$ erg/s in NUV band.
 
\begin{figure*}
\centering
\begin{center}
\includegraphics[angle=0, width=\textwidth]{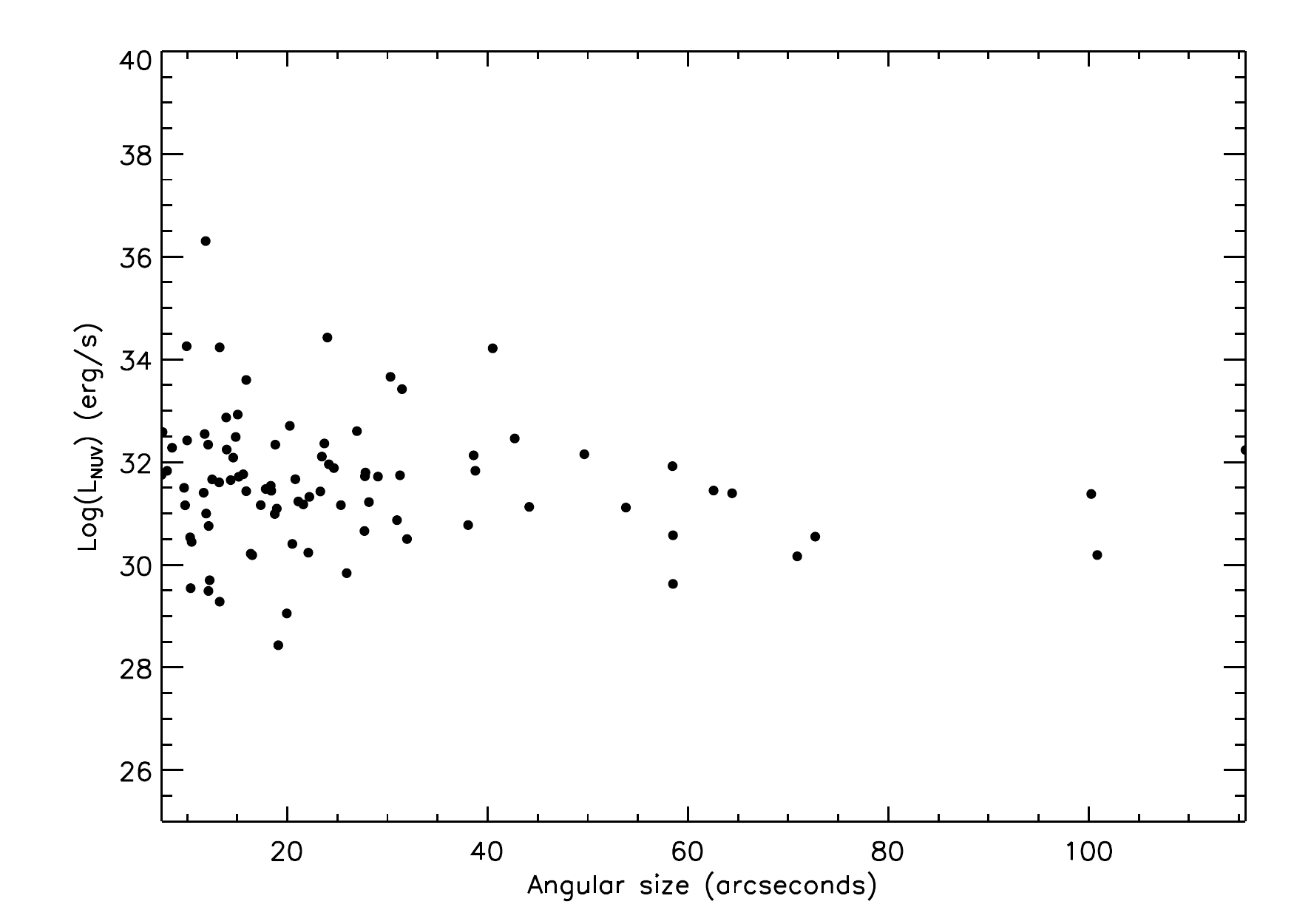}
\caption{The NUV luminosity is plotted against the NUV angular size for 89 PNe. For the non-spherical PNe, the average values of the major and minor axes are used.}
\label{fig3}
\end{center}
\end{figure*}

 It is seen from Table \ref{tab1} that the L$_{\text{FUV}}$ is higher than the L$_{\text{NUV}}$ when we compare the luminosities of the PNe for which observation in both the bands exists. \citet{Weid11} have confirmed with their spectroscopic analysis that most of the central stars of PNe are OB type and emission line stars. The CSPNe temperature ranges from 40 to 120 kK and emit bulk of their fluxes in UV. When we compare the luminosity in two filters of the GALEX, it shows that the L$_{\text{FUV}}$ is
higher than the L$_{\text{NUV}}$ despite the FUV filter having narrower bandwidth than the NUV filter.  Most of the UV radiation of the PNe is dominated by its central star and the higher value of L$_{\text{FUV}}$ indicates that the CSPNe are brighter in FUV than NUV. In Figure \ref{fig3}, we plot NUV angular size (column 9 of Table \ref{tab1}) versus NUV luminosity (column 6 of Table \ref{tab1}) for 89 PNe. Both these quantities are measured from the \textit{GALEX} images. The Figure \ref{fig3} shows that the L$_{\text{NUV}}$ of the PNe falls in a relatively narrow range and does not vary much with the angular size. Again, this indicates that the contribution of the CSPNe is much more than the nebular continuum emission to the total integrated flux of the PNe. Hence, with increase in angular size the luminosity of PNe does not increase much. 

\begin{figure*}
    \centering
    \includegraphics[width=\textwidth]{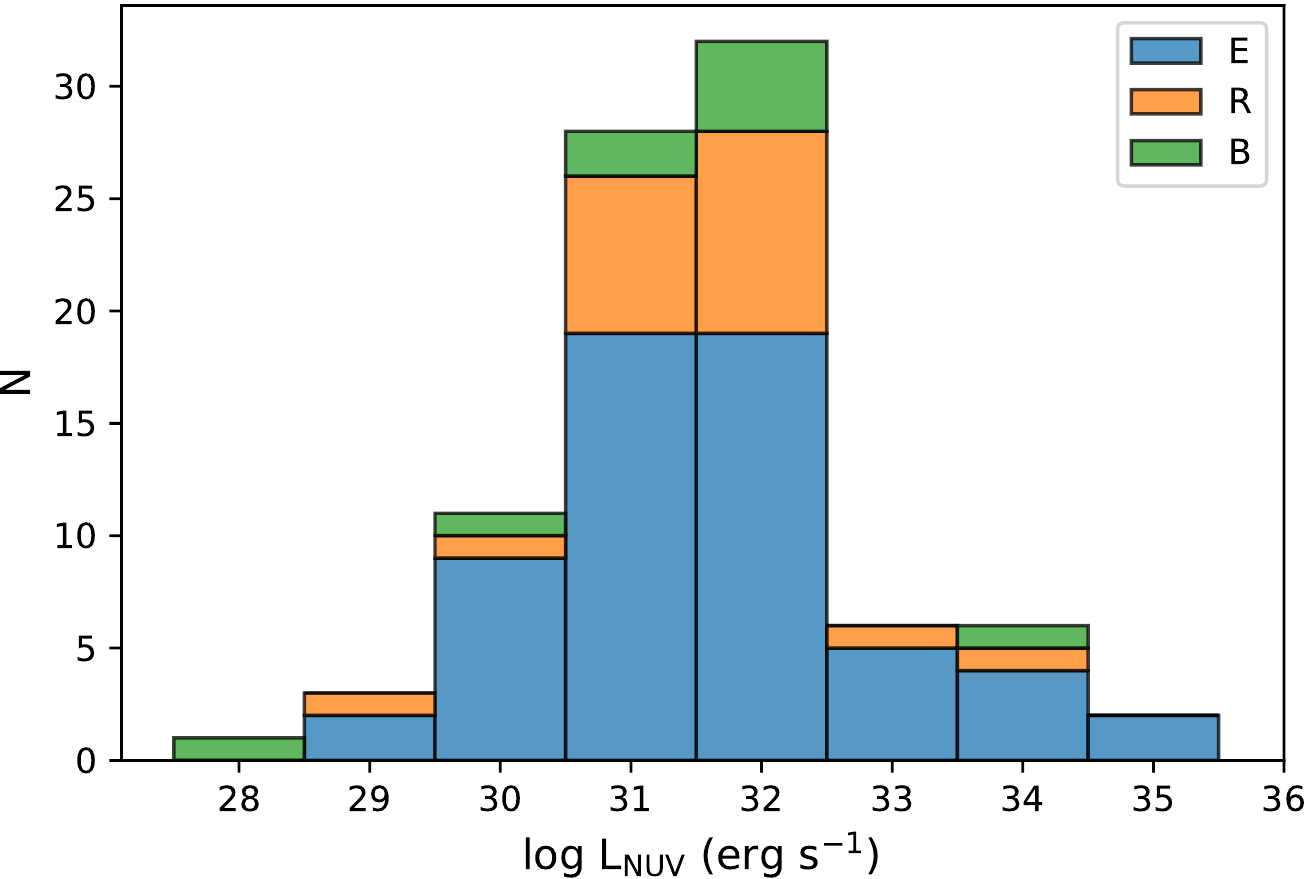}
    \caption{Stacked histogram showing the distribution of NUV luminosity for the 89 PNe. The colors denote the three morphological classes - elliptical (E), round (R) and bipolar (B).}
    \label{figx}
\end{figure*}

Similar to Figure 1 in \citep{Weid11}, we show the distribution of the \textit{GALEX} NUV luminosities as a function of the morphological classes -- elliptical (E), round (R) and bipolar (B) in Figure \ref{figx}. We see that in our sample the elliptical-types are twice as populated compared to the round type and thrice as compared to the bipolar type. Also, the ellipticals span in a large range in L$_{\text{NUV}}$ as compared to the other two types.
 
\begin{figure*}
    \centering
    \includegraphics[width=0.75\textwidth]{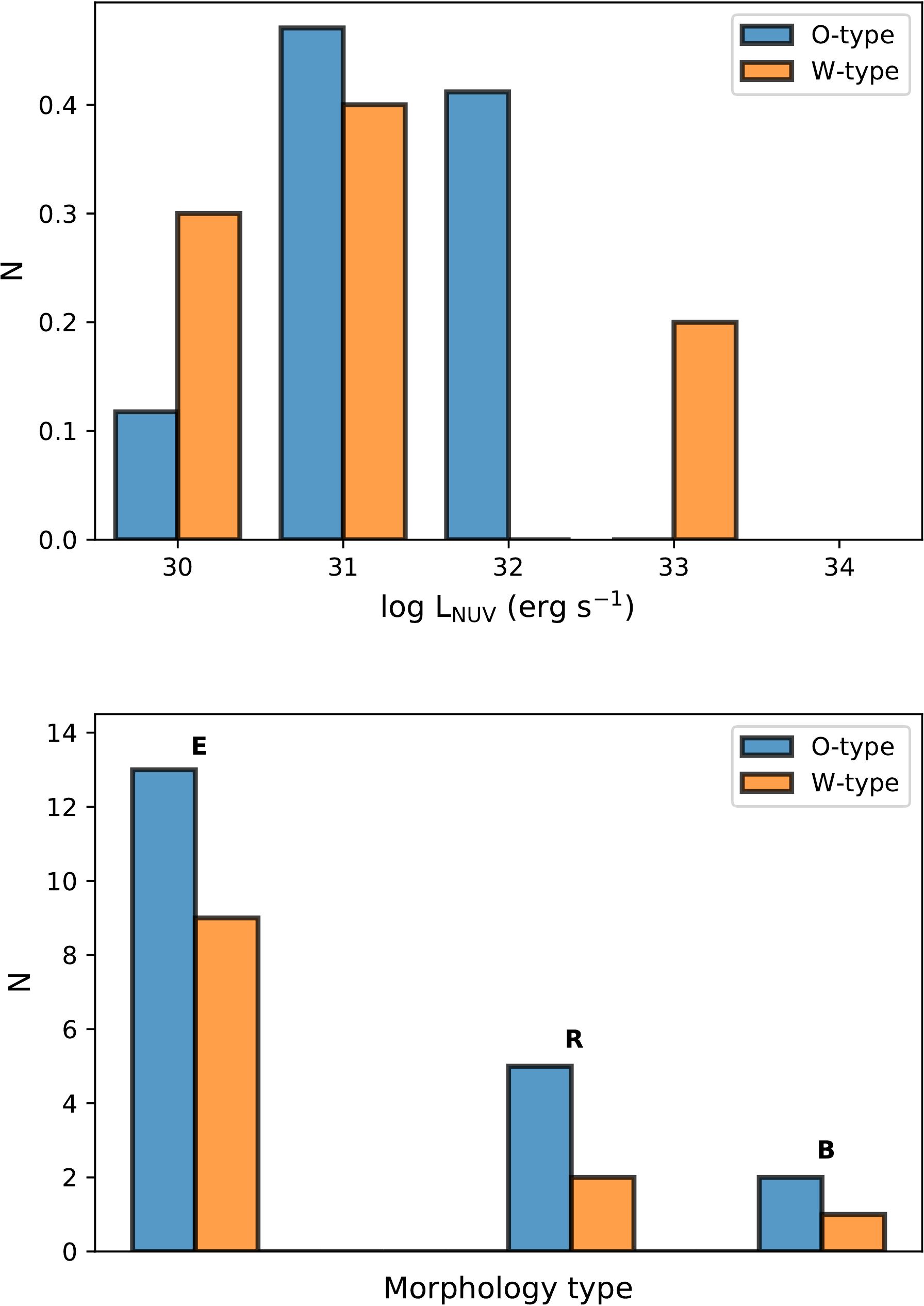}
    \caption{Histograms showing the distribution of (a) NUV luminosity and (b) the type of morphology for two subsamples -- O-type (N = 20) and W-type (N = 12). The acronyms for the three morphological classes are same as in Fig. 4.}
    \label{figy}
\end{figure*}

We divide our sample into two sub-samples, one containing the nebulae with core [WR] and another containing PNe with nuclei O-type to study if there is any difference between the parameters of both the sub-samples, especially their UV morphology and NUV luminosity (see Figure \ref{figy}). In general, we do not find any major difference between the two populations. We have an uneven sample with a relatively higher number of sources with elliptical morphology than the other two morphological classes.  In all the three kinds of morphology (elliptical, round and bipolar) there are more sources with O-type than W-type (in elliptical: by 19\%, in round: by 43\%, in bipolar: by 33\%). The W-type have a large spread in the NUV luminosities. On the other hand O-type are more localized between L$_{\text{NUV}}$ = 10$^{30}$ - 10$^{33}$ ergs s$^{-1}$.
 
  It is also anticipated that the stellar wind of the sources with high value of FUV flux may produce X-ray emission in its proximity. \citet{Feld97} theoretically predicts that instabilities in the line-driven stellar winds from CSPNe may also produce X-ray emitting shocks in immediate vicinity of the central star (O-stars). It is also demonstrated that the X-ray fluxes from shocks are necessary to explain the UV emission from high ionization species of oxygen (O VI and O VII) in the coolest CSPNe and can also play a role at higher CSPN temperatures \citep{Herald11, Kas12, Guerrero13}. Recently there have been several observations of PNe by \textit{Chandra} X-ray observatory and  \textit{XMM-Newton} which have discovered both X-ray diffuse and point like observations \citep{Sok03, Kastner12, Freeman14, Montez15}. We found six X-ray detections in our sample (PN G066.7$-$28.2, PN G083.5$+$12.7, PN G096.4$+$29.9, \& PN G165.5$-$15.2 from \citet{Montez15}, and PN G084.9$-$03.4, and PN G226.7$+$05.6 from \citet{Sok03}), which contain point-like sources. The X-ray luminosities for the sources, that are likely to be coming from the CSPNe are measured. We found that the derived L$_{\text{FUV/NUV}}$ are $1-3$ orders more than the corresponding X-ray luminosities. The higher value of UV luminosity than the X-ray might be due to the fact the UV emission is integrated emission coming from the entire extended PN as compared to the very compact CSPNe that gives rise to the bulk of X-ray emission.

 \subsection{\textit{IUE} spectra of PNe and \textit{GALEX} filters}

\begin{figure*}
  \centering
   \includegraphics[height=2.50in,width=5in, trim=0cm 0.0cm 0cm 0cm, clip=true, angle=0]{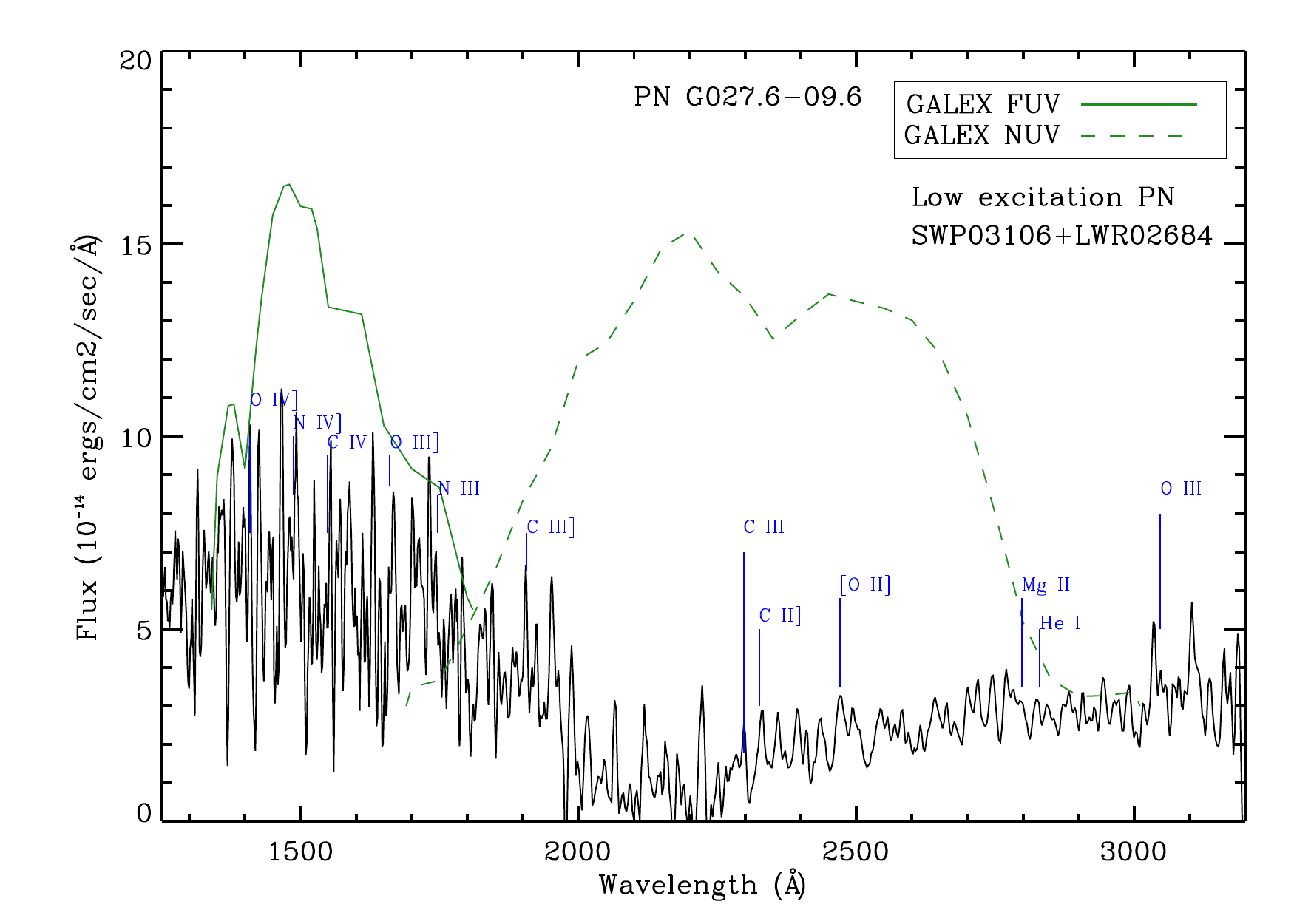}
   \includegraphics[height=2.50in,width=5in, trim=0cm 0.0cm 0cm 0cm, clip=true, angle=0]{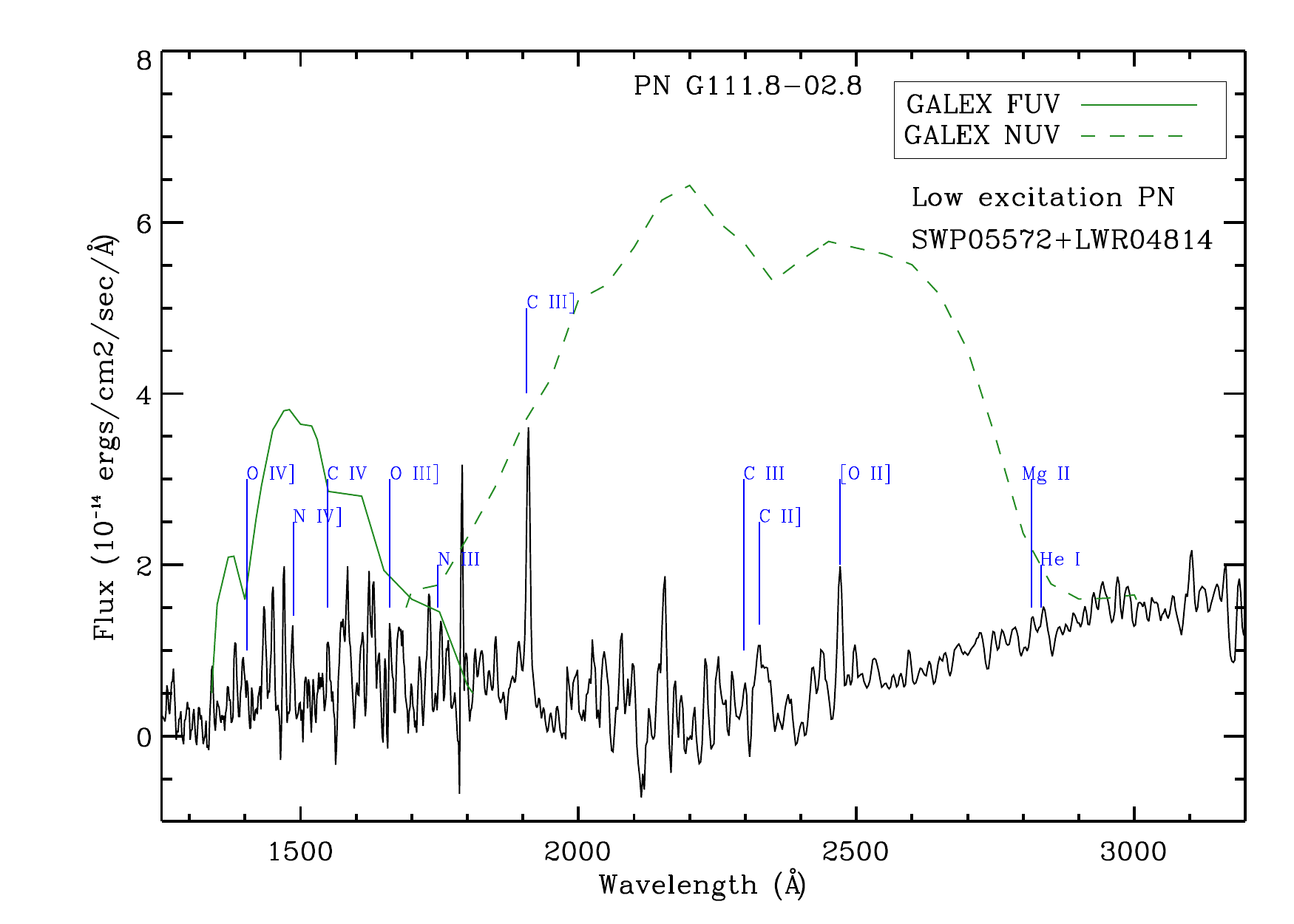}
   \includegraphics[height=2.50in,width=5in, trim=0cm 0.0cm 0cm 0cm, clip=true, angle=0]{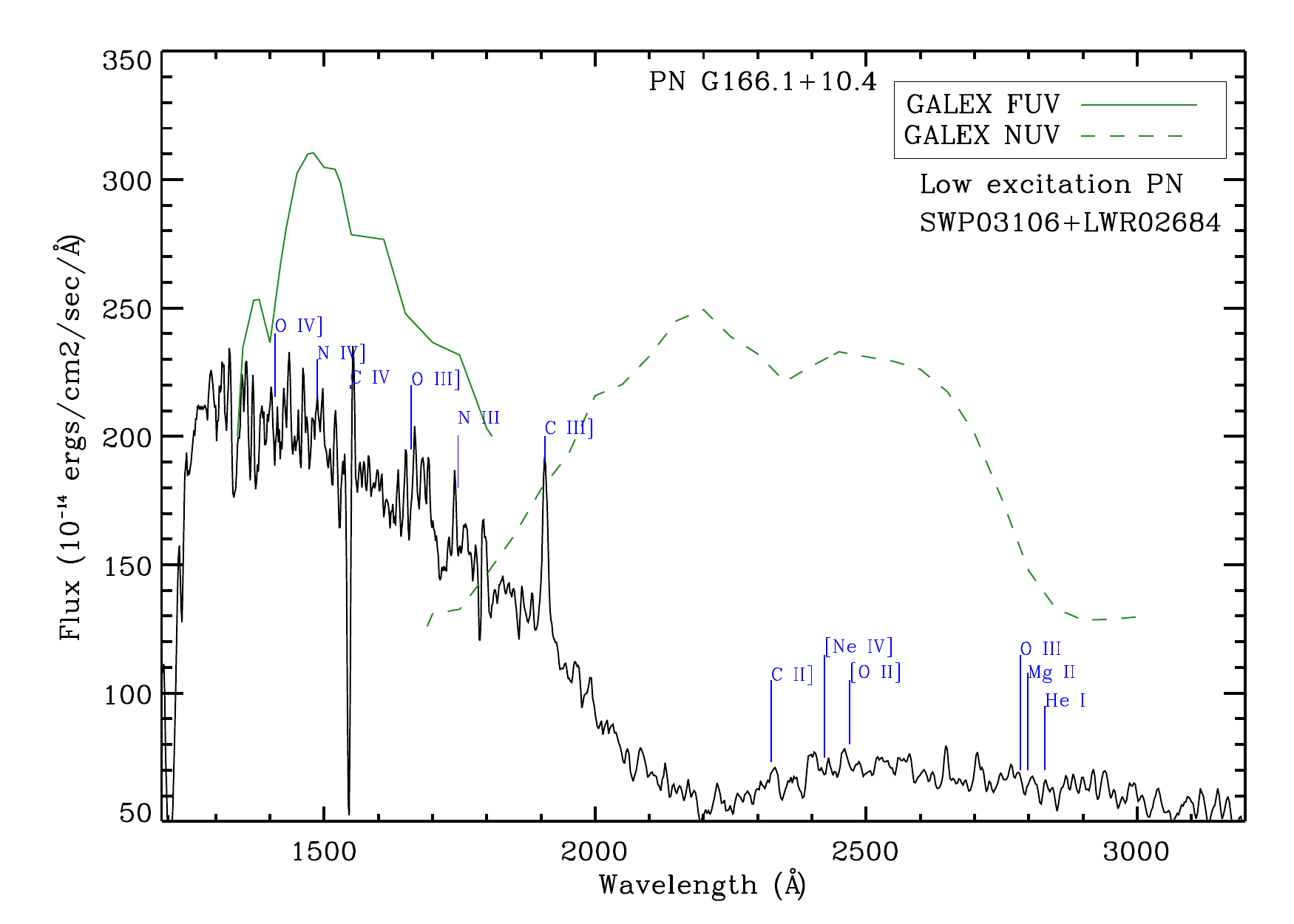}
   \caption{\textit{IUE} spectra of three low excitation PNe (PN G027.6-09.6, PN G111.8$-$02.8, and PN G166.1$+$10.4) which are smoothen by three points. The strong emission lines are marked. The green solid and dashed lines are the effective area curves of \textit{GALEX} FUV and NUV filters, respectively.}
\label{fig4}
\end{figure*}

\begin{figure*}
\centering
\includegraphics[height=2.50in,width=5in, trim=0cm 0.0cm 0cm 0cm, clip=true, angle=0]{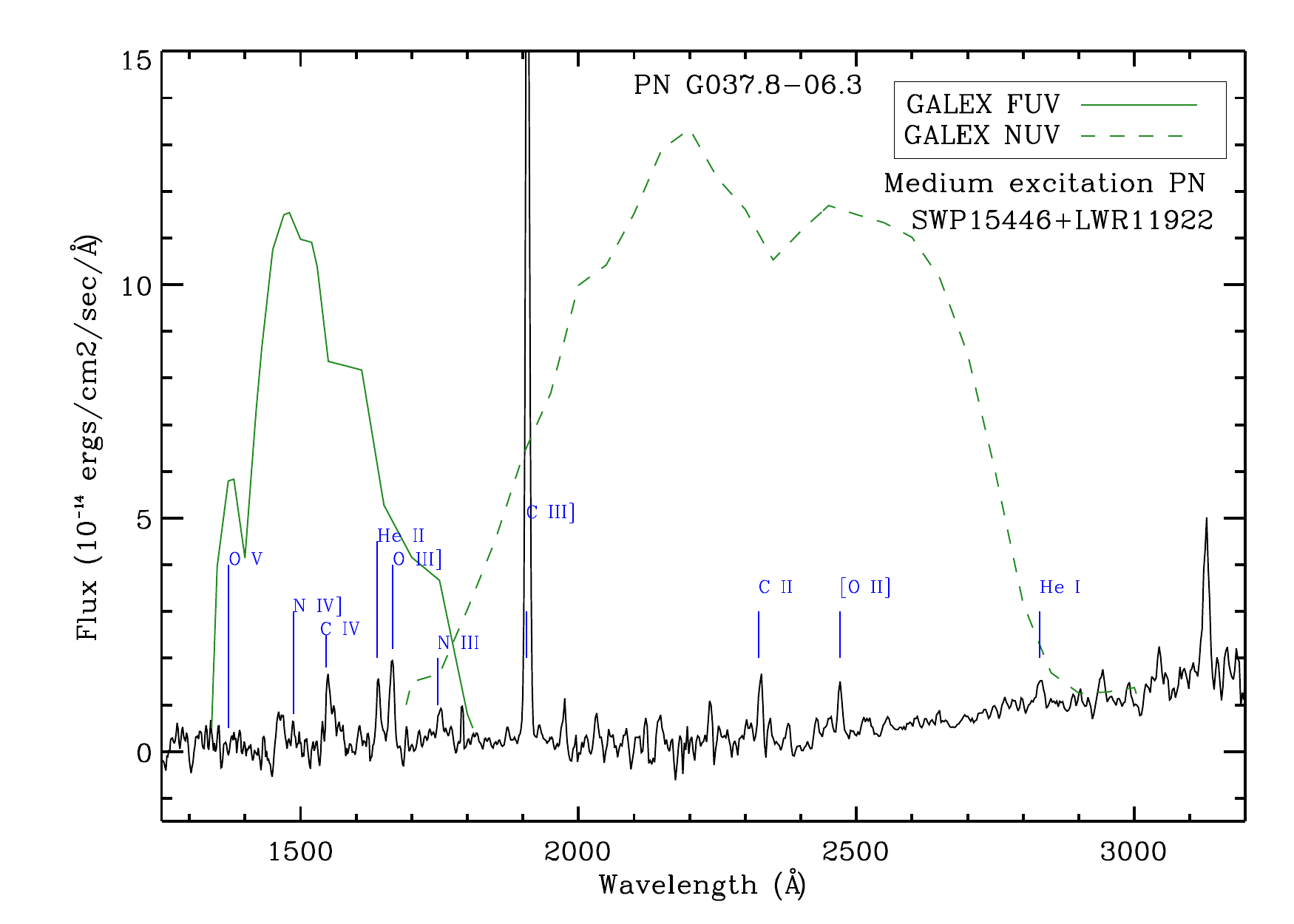}
   \includegraphics[height=2.50in,width=5in, trim=0cm 0.0cm 0cm 0cm, clip=true, angle=0]{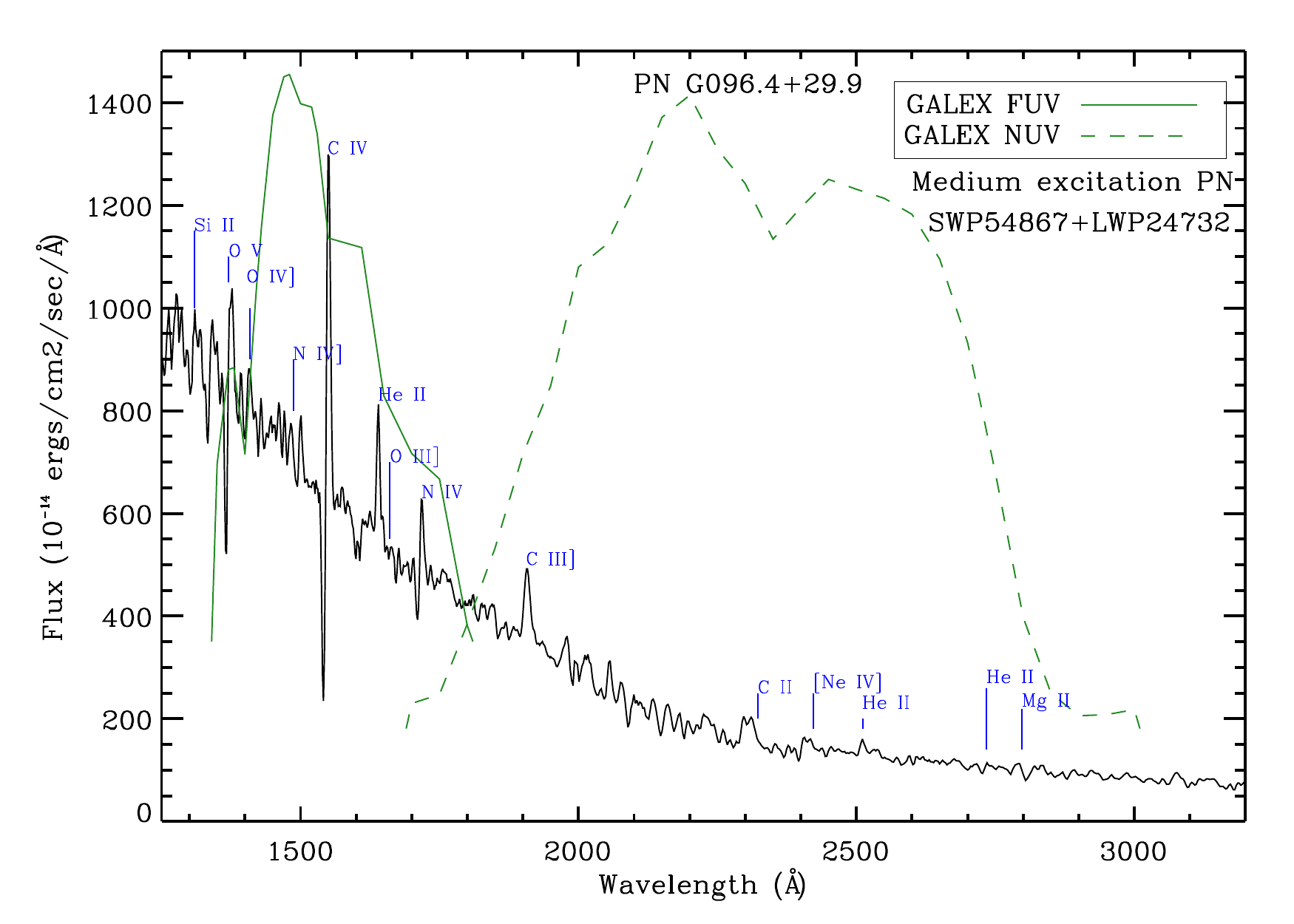}
  \includegraphics[height=2.50in,width=5in, trim=0cm 0.0cm 0cm 0cm, clip=true, angle=0]{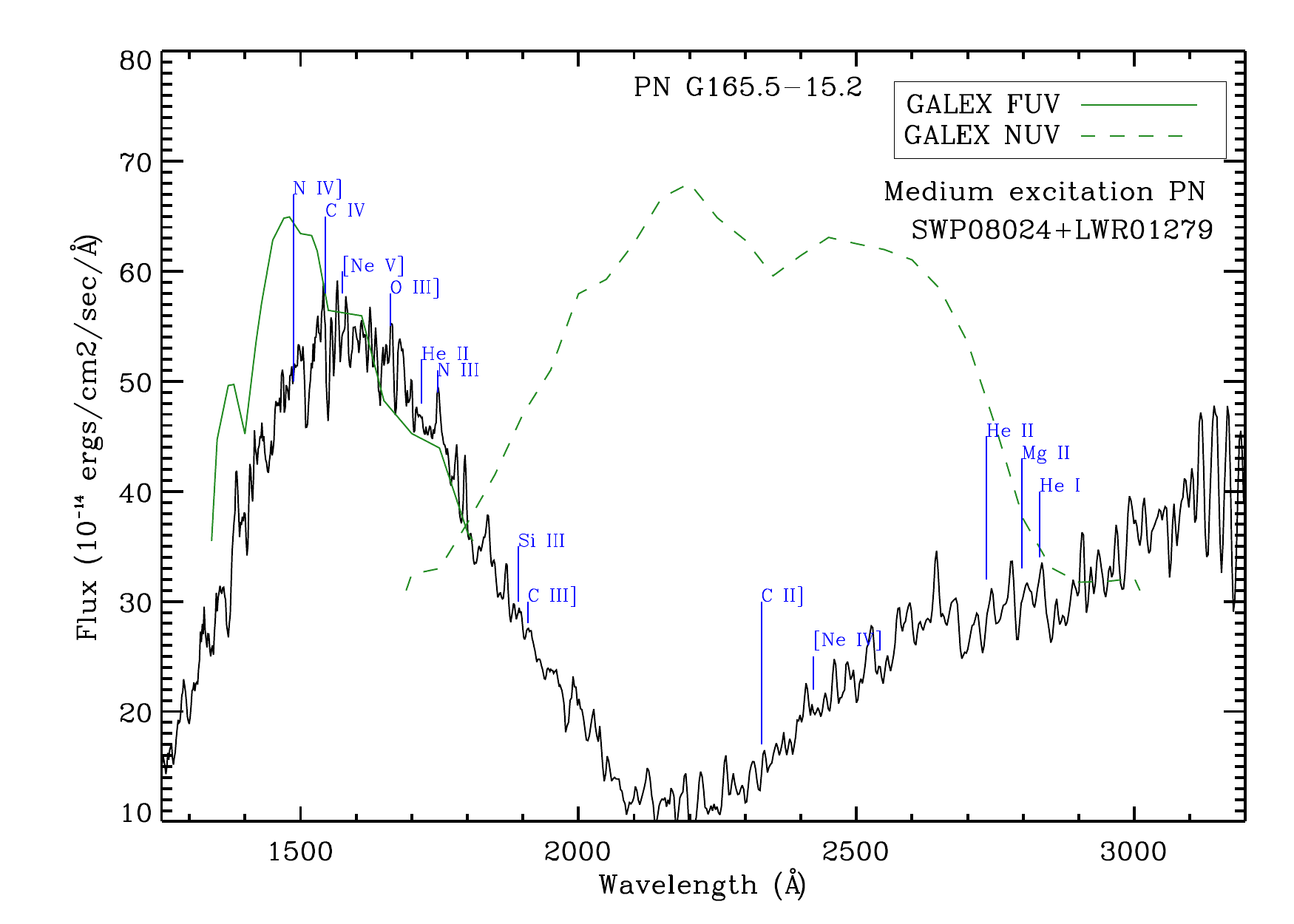}
\caption{\textit{IUE} spectra of three medium excitation PNe (PN G037.8$-$06.3, PN G096.4$+$29.9, and PN G165.5$-$15.2 ) which are smoothen by three points. The strong emission lines are marked. The green solid and dashed lines are the effective area curves of \textit{GALEX} FUV and NUV filters, respectively.}
\label{fig5}
\end{figure*}

\begin{figure*}
\centering
\includegraphics[height=2.50in,width=5in, trim=0cm 0.0cm 0cm 0cm, clip=true, angle=0]{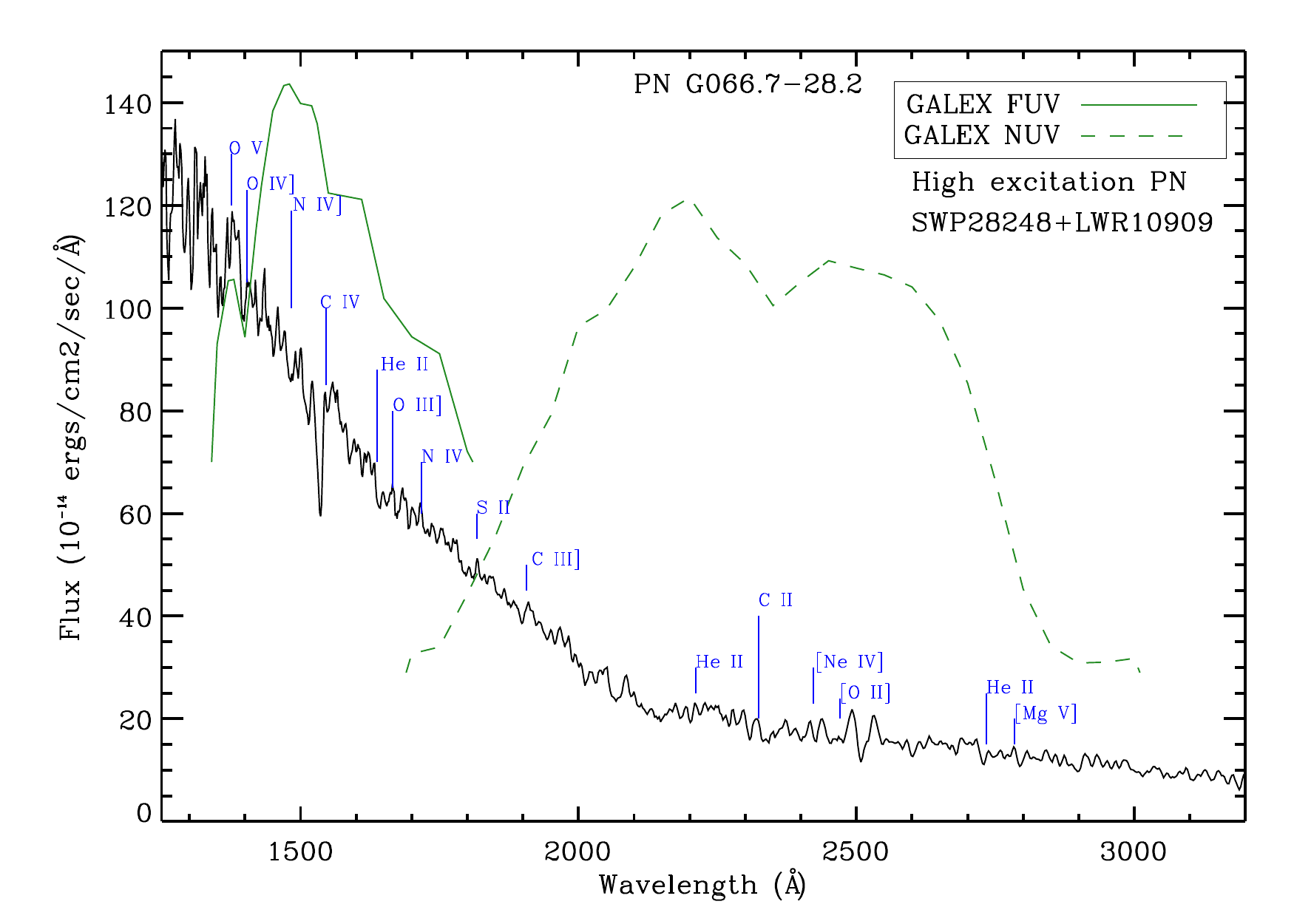}
   \includegraphics[height=2.50in,width=5in, trim=0cm 0.0cm 0cm 0cm, clip=true, angle=0]{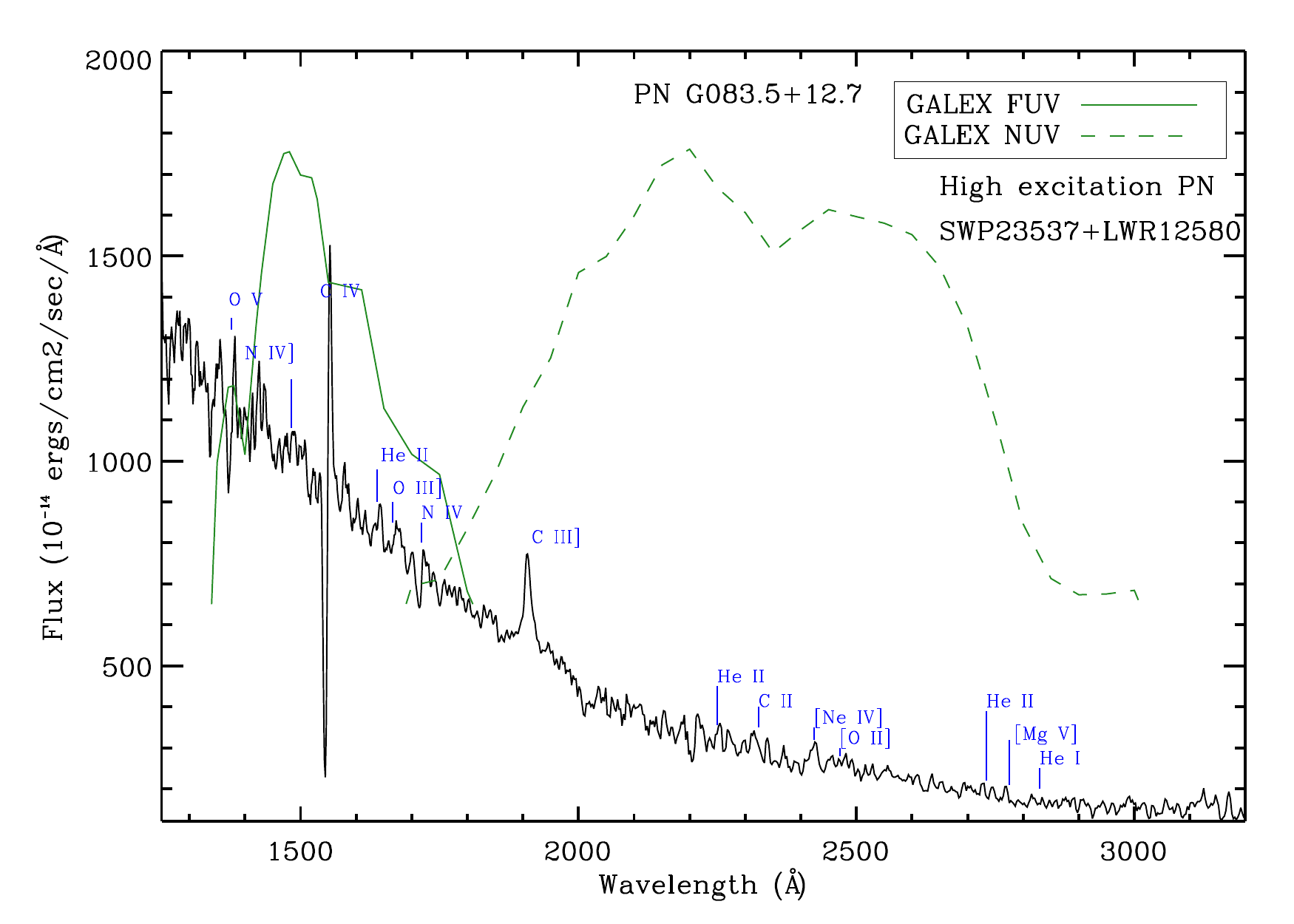}
   \includegraphics[height=2.50in,width=5in, trim=0cm 0.0cm 0cm 0cm, clip=true, angle=0]{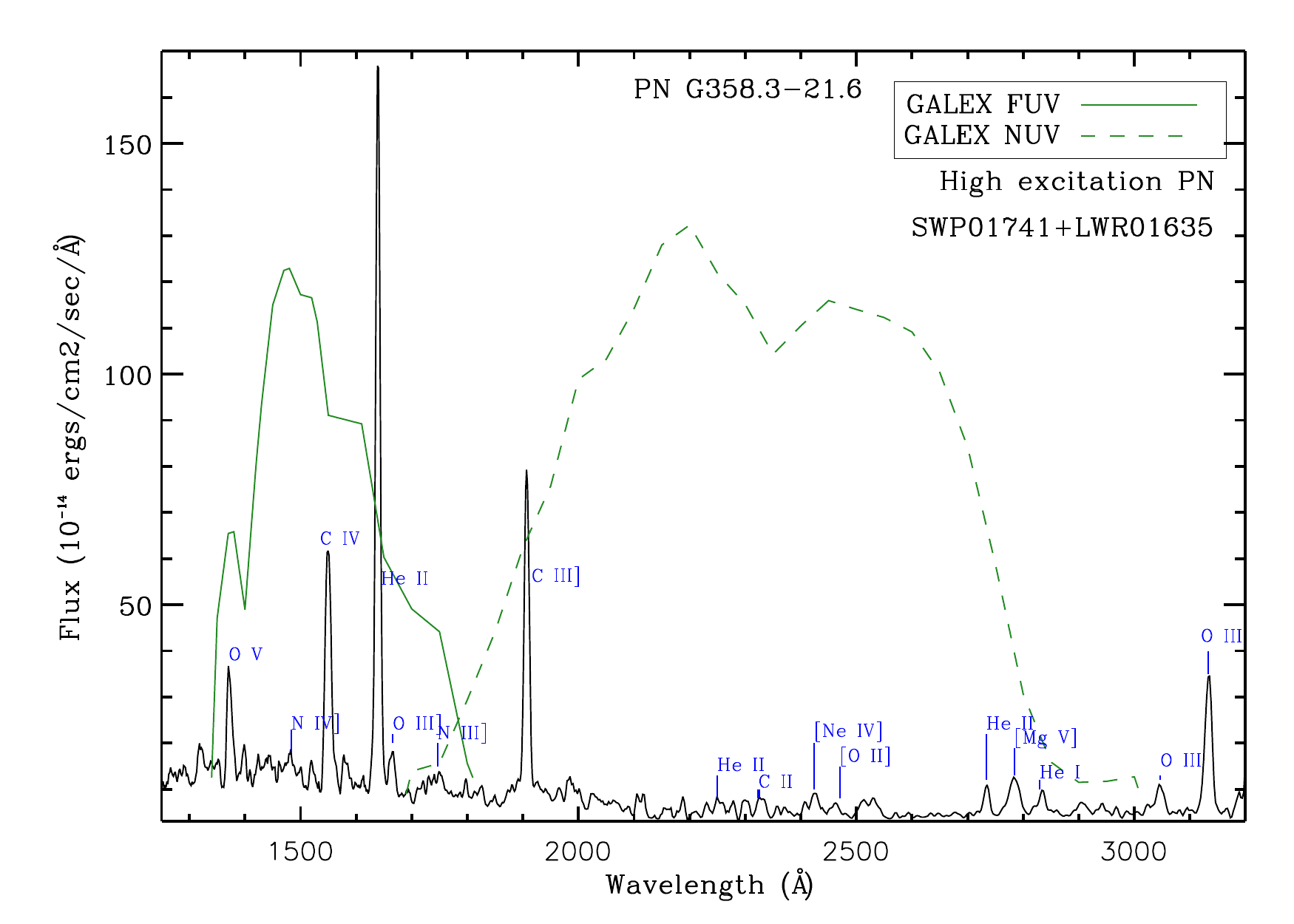}
\caption{\textit{IUE} spectra of three high excitation PNe (PN G066.7$-$28.2, PN G083.5$+$12.7 and PN G358.3$-$21.6) which are smoothen by three points. The strong emission lines are marked. The green solid and dashed lines are the effective area curves of \textit{GALEX} FUV and NUV filters, respectively.}
\label{fig6}
\end{figure*}

The UV analysis of CSPNe has been made for about 300 PNe using the spectra obtained by \textit{IUE} \citep{Heap87, Koppen87, Heap97, Pauldrach04, Herald11, Keller14}. We found several of the PNe detection in our \textit{GALEX} sample have \textit{IUE}\footnote{\href{https://archive.stsci.edu/iue/}{https://archive.stsci.edu/iue/}} spectra which are useful in explaining the nature of UV emission in the \textit{GALEX} bands. The \textit{GALEX} FUV band is completely contained in \textit{IUE} SWP band (1150 - 1980 \AA) whereas NUV band is split between the LWR (1850 - 3300 \AA) and SWP of \textit{IUE}. In order to compare the \textit{GALEX} NUV images with \textit{IUE} spectra of PNe, the SWP and LWR spectra are joined into a single spectrum setting the matching wavelength at 1975 \AA. Instead of showing all the 42 \textit{IUE} spectra available for our PNe sample, we have shown nine spectra three each from low, medium and high excitation PNe (Figure \ref{fig4}, \ref{fig5} and \ref{fig6}, respectively). The information regarding the excitation class of the PNe is taken from \citet{Gur91}. Since \textit{GALEX} and \textit{IUE} have almost similar wavelength coverage, we have displayed the filter response curves of \textit{GALEX} FUV and NUV filters on each of the combined \textit{IUE} spectra.  We observed numerous UV emission lines in the \textit{IUE} spectra of PNe. The list of emission lines which contribute to the UV fluxes in \textit{GALEX} bands are categorically mentioned for the PNe of different excitation classes as follows. \\
\textbf{1. Low excitation PNe (Figure \ref{fig4}):} The emission lines that contribute to the flux of \textit{GALEX} FUV band are O IV] (1403 \AA), N IV] (1487 \AA), C IV (1550 \AA), O III] (1661 \AA) and N III (1892 \AA) whereas the lines that contribute to the flux of \textit{GALEX} NUV band are C III] (1907 \AA) and C II (2325 \AA), O II (2470 \AA), Mg II (2830 \AA) and He I (2830 \AA). The C IV and C III] lines are the strongest lines in FUV and NUV, respectively, which might be contributing more  than the other lines in the respective bands. However, there are many other lines that contribute to the fluxes in both the bands but they are not common in all the three PNe as shown in Figure \ref{fig4}.\\
\textbf{2. Medium excitation PNe (Figure \ref{fig5}):} The emission lines that contribute to the flux of \textit{GALEX} FUV band are N IV] (1487 \AA), C IV (1550 \AA), He II (1661 \AA) and O III] (1661 \AA) whereas the lines that contribute to the flux of NUV band are C III] (1907 \AA) and C II (2325 \AA). There are many other lines that contribute to the fluxes in both the bands but all these lines they are not common in all the three PNe as shown in Figure \ref{fig5}.\\
\textbf{3. High excitation PNe (Figure \ref{fig6}):}  The emission lines that contribute to the flux of FUV band are N IV] (1487 \AA), C IV (1550 \AA), He II (1661 \AA) and O III] (1661 \AA) whereas the lines that contribute to the flux of NUV band are C III] (1907 \AA), He II (2252 \AA), [O II] (2470 \AA), He II (2734 \AA) and [Mg V] (2784 \AA). Similar to the two cases mentioned before, there are many other emission lines which are not common in all three spectra but they do contribute to the fluxes in both the bands. \\

However, there are dissimilarities even  among the spectra of PNe of similar excitations in the sense that some of the emission lines are not common for all. So, the contribution of a particular emission line to the nebular emission of PNe of different excitation classes may or may not be same and depends on the individual PNe. Again the diameter of most of the PNe in our sample are larger than the large entrance aperture of the low resolution \textit{IUE} observations, therefore, the \textit{IUE} spectra include the central star and a small part of the nebula. Whereas, the \textit{GALEX} images include the central star and nebular continuum along with nebular emission lines.

\subsection{\textit{GALEX} images of selected PNe}

\begin{figure*}
\centering
\includegraphics[width=15.5cm]{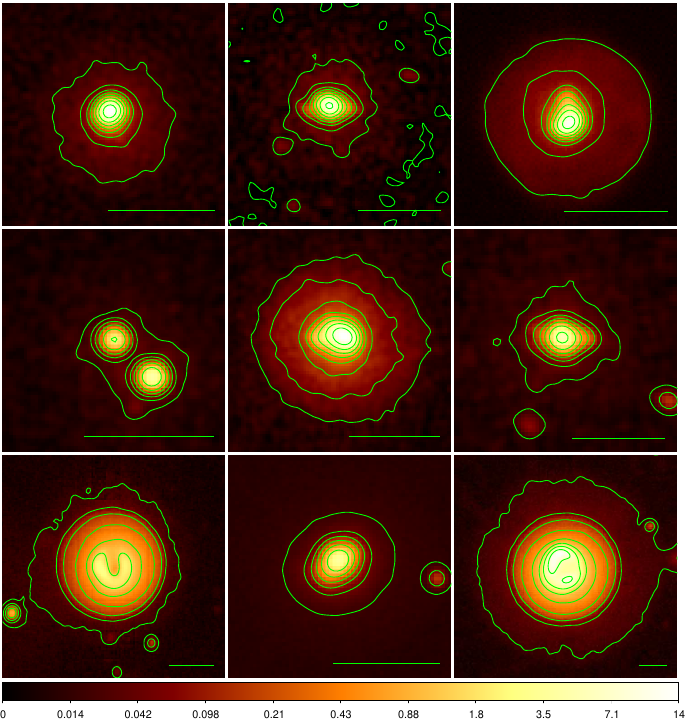}
\caption{NUV images of selected PNe with NUV contours (10 levels and smoothness parameter 2). From left to right, top to bottom: (1) PN G013.3$+$32.7; (2) PN G061.9$+$41.3; (3) PN G066.7$-$28.2; (4) PN G072.7$-$17.1; (5) PN G081.2$-$14.9; (6) PN G082.5$+$11.3; (7) PN G083.5$+$12.7; (8) PN G084.9$-$03.4; (9) PN G096.4$+$29.9. The scale bars below are $4''$.}
\label{fig7a}
\end{figure*}

\begin{figure*}
\centering
\includegraphics[width=15.5cm]{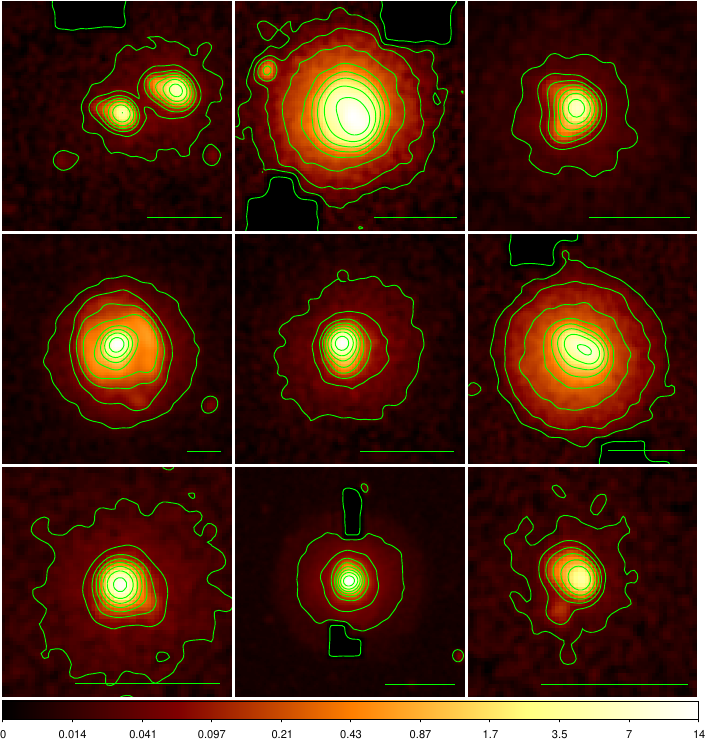}
\caption{NUV images of selected PNe with NUV contours (10 levels and smoothness parameter 2). From left to right, top to bottom: (10) PN G100.0$-$08.7; (11) PN G106.5$-$17.6; (12) PN G107.6$-$13.3; (13) PN G120.0$+$09.8; (14) PN G165.5$-$15.2; (15) PN G166.1$+$10.4; (16) PN G190.3$-$17.7; (17) PN G208.5$+$33.2; (18) PN G222.1$+$03.9. The scale bars below are $4''$.}
\label{fig7b}
\end{figure*}

\begin{figure*}
\centering
\includegraphics[width=15.5cm]{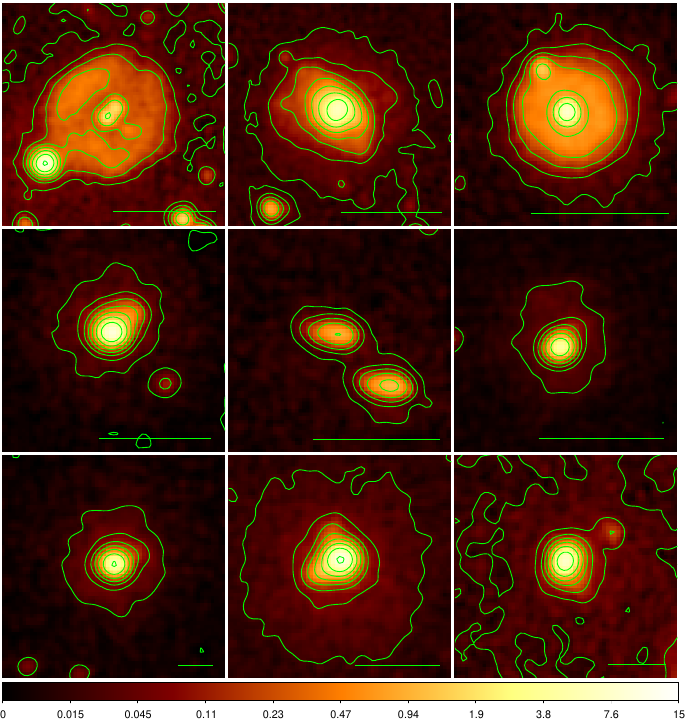}
\caption{NUV images of selected PNe with NUV contours (10 levels and smoothness parameter 2). From left to right, top to bottom: (19) PN G231.8$+$04.1; (20) PN G234.8$+$02.4; (21) PN G239.6$+$13.9; (22) PN G243.8$-$37.1; (23) PN G249.0$+$06.9; (24) PN G270.1$+$24.8; (25) PN G286.8$-$29.5; (26) PN G291.3$+$08.4; (27) PN G309.0$-$04.2. The scale bars below are $4''$.}
\label{fig7c}
\end{figure*}

\begin{figure*}
\centering
\includegraphics[width=15.5cm]{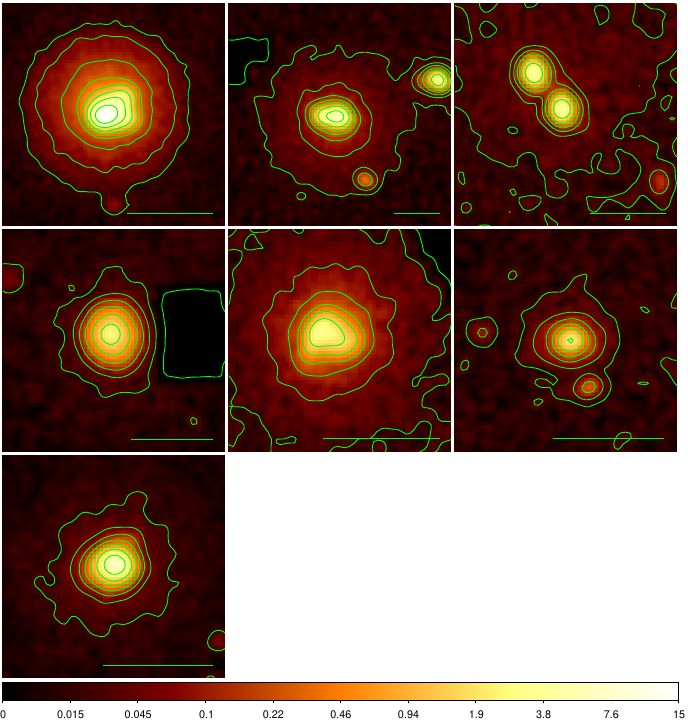}
\caption{NUV images of selected PNe with NUV contours (10 levels and smoothness parameter 2). From left to right, top to bottom: (28) PN G310.3$+$24.7; (29) PN G325.8$-$12.8; (30) PN G327.7$-$05.4; (31) PN G334.3$-$09.3; (32) PN G341.6$+$13.7; (33) PN G353.7$-$12.8; and (34) PN G358.3$-$21.6. The scale bars below are $4''$.}
\label{fig7d}
\end{figure*}

\begin{figure*}
\centering
\includegraphics[width=15.5cm]{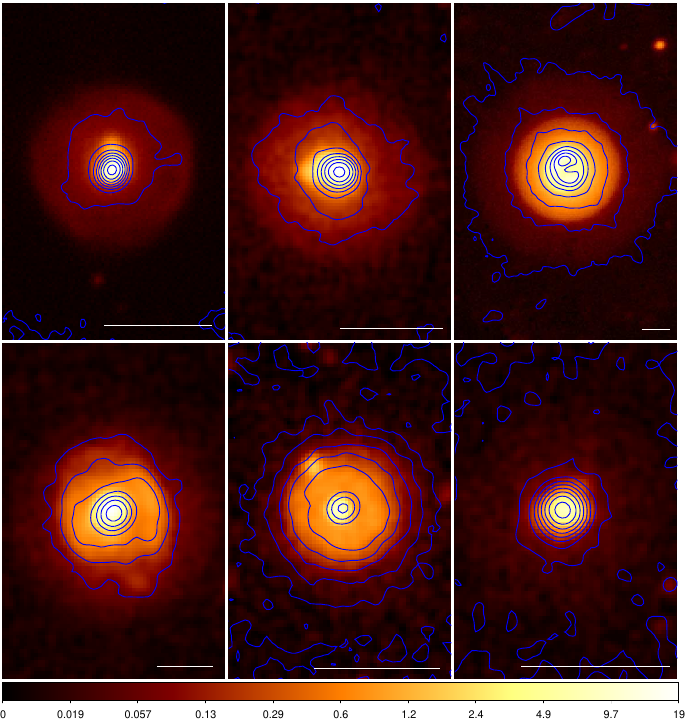}
\caption{NUV images of selected PNe with FUV contours (10 levels and smoothness parameter 2). From left to right, top to bottom: (1) PN G066.7$-$28.2; (2) PN G081.2$-$14.9; (3) PN G096.4$+$29.9; (4) PN G120.0$+$09.8; (5) PN G239.6$+$13.9; and (6) PN G358.3$-$21.6. The scale bars below are $4''$.}
\label{fig8}
\end{figure*}

\begin{figure*}
\centering
\includegraphics[width=10cm]{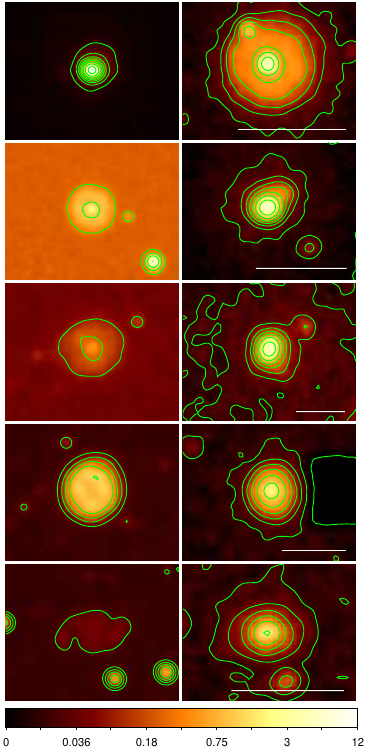}
\caption{Comparing common PNe images in [N II] (\citealt{Weid16}; on the left) and NUV (this paper; on the right) with respective contours (from top to bottom): (1) PN G239.6$+$13.9; (2) PN G243.8$-$37.1; (3) PN G309.0$-$04.2; (4) PN G334.3$-$09.3; and (5) PN G353.7$-$12.8. The scale bars below are $4''$ only shown for the NUV images.}
\label{fig9}
\end{figure*}

In the sample of 108 extended PNe, we have provided contour images of 34 PNe in NUV and in FUV (Figures \ref{fig7a}, \ref{fig7b}, \ref{fig7c}, \ref{fig7d} and Figure \ref{fig8}) which show rich structures. The contour maps at both the bands are in logarithmic scale of intensity. The central levels are at the highest value of intensity with the trend decreasing towards the external contours. The various contour levels show the structures of PNe in UV.  The near-concentric contours in the case of FUV relative to that of its NUV counterpart is the characteristic of the dominating flux - the extension at the two wavelength regimes, wherein the peak in the intensity value in the FUV can be attributed to the relative hotness of the CSPNe. Recently, \citet{Weid16} have provided the [N II] narrow band images of 108 PNe along with their brief morphology. Eight sources (PN G014.8$-$25.6, PN G226.7$+$05.6, PN G 239.6$+$13.9, PN G 243.8$-$37.1, PN G249.8$+$07.1, PN G 309.0$-$04.2, PN G 334.3$-$09.3, and PN G 353.7$-$12.8) from their catalog match with our sample. We show the comparison for five of the eight sources with NUV morphology in Figure \ref{fig9}. The rich morphological features of the PNe in \textit{GALEX} NUV band are nicely discernible as compared to the [N II] narrow band images.
\par
\begin{figure*}
\centering
\includegraphics[width=15.5cm]{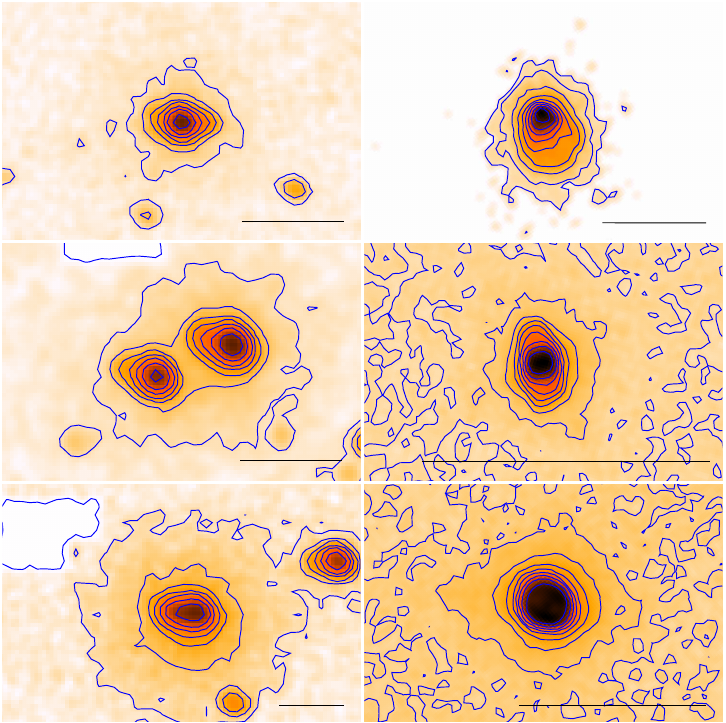}
\caption{Comparing common PNe images in NUV (this paper; on the left) and \textit{HST} (\citealt{Sahai11}; on the right) with respective contours (from top to bottom): (1) PN G082.5$+$11.3; (2) PN G100.0$-$08.7; and (3) PN G325.8$-$12.8. The scale bars below are $4''$ for the NUV images and $1''$ for the \textit{HST} images.}
\label{fig10}
\end{figure*}

 
We performed a similar comparison for \textit{HST} images from \cite{Sahai11}. Among the 22 sources found common, only 3 showed significant level of details in their contour images to be compared - PN G082.5$+$11.3, PN G100.0$-$08.7 and PN G325.8$-$12.8. (see Figure \ref{fig10}). The first and the third sources were found to belong to elliptical class, but the second source showed dramatically different morphology (bipolar instead of elliptical, as per the corresponding \textit{HST} image.

\section{Conclusions}

We have identified about 358 PNe detected by \textit{GALEX} observation and provided a catalog of angular sizes  in NUV for 108 and in FUV for 28 of them. A comparative study of angular sizes against effective wavelengths in four distinct wavelength bands is given for 24 extended PNe. Using axial-ratio method, we have discussed morphological classification of the PNe in NUV. Depending on the availability of distance to the PNe, the intrinsic luminosities are calaculated for 89 PNe in NUV and for 33 PNe in FUV. We also found that the L$_{\text{NUV}}$ of the PNe falls in a relatively narrow range and does not vary much with the angular size suggesting that the dominant contribution to the integrated UV flux of the PNe comes from the central star. The higher value of L$_{\text{FUV}}$ than the L$_{\text{NUV}}$ indicates that the PNe are brighter in FUV than NUV. We did not find any difference in parameters such as luminosity, morphology and size between two sub-samples, one containing the nebulae with core [WR] and another containing PNe with nuclei O-type. We have presented  contour images for 34 PNe in NUV and for 9 PNe, their respective FUV contours are over-plotted over the corresponding NUV images. We anticipate that these images and size information in UV will provide clue to future research in PNe.

The nebular emission in the \textit{GALEX} FUV band is attributed to prominent emission lines N IV] (1487 \AA), C IV (1550 \AA), and O III] (1661 \AA) and the emission in the \textit{GALEX} NUV band is attributed to C III] (1907 \AA) and C II (2325 \AA) for PNe of all excitation classes. Apart from these lines, there are many other emission lines that contribute to the FUV and NUV in all excitation classes.
However, it is not possible to disentangle the central star UV continuum and contribution from the nebular emission lines because of low resolution of the \textit{GALEX} images. Also, the accounting of the different contributions to the net integrated fluxes of the PNe detected in \textit{GALEX} requires proper ionization modeling which is out of the scope of this paper.

\section*{Acknowledgments}

 We thank Prof. W. A. Weidmann for his valuable and helpful comments.
GALEX (Galaxy Evolution Explorer) is a NASA Small Explorer, launched
in April 2003. We gratefully acknowledge NASA's support for construction, operation, and science analysis for the \textit{GALEX} mission, developed in cooperation with the Centre National d'Etudes Spatiales (CNES) of France and the Korean Ministry of Science and Technology. Author ACP acknowledges financial support received under Fast Track Scheme for Young Scientist SERB-DST, New Delhi, India (File Number: YSS/2015/000114). ACP also acknowledges IUCAA, Pune for providing facilities to carry out the research. SP would like to acknowledge Polish Funding Agency National Science Centre, projects 2015/17/B/ST9/03436/ (OPUS 9) and 2017/26/A/ST9/00756 (MAESTRO  9). MP is thankful to Prof. Wako Aoki and  Director General of National Astronomical Observatories of Japan (NAOJ) Prof. Saku Tsuneta for their kind encouragement and support.

\input{table3.tex}
\input{table4.tex}

\clearpage


\end{document}

%% file: table3.tex

\begin{landscape}
\begin{table}
\caption{Summary of the 108 extended PNe detected by \textit{GALEX}}
\footnote{aaa}
\label{tab1} 
\scriptsize{
\begin{tabular}{ccccccccccccc}
\hline
PN G & Name & RA (J2000) & Dec (J2000) & E(B-V) & L\_NUV & L\_FUV & ST of CSPN & NUV\_size & FUV\_size & V\_size & H$\alpha$\_size & type \\
 &  & (hh mm ss) & (hh mm ss) & & (erg s$^{-1}$) & (erg s$^{-1}$) &  & (arcsec) & (arcsec) & (arcsec) & (arcsec) &  \\
\hline

001.0$+$01.9 & K 1$-$4        & 17 40 28.04 & $-$27 00 48.25  & 2.067  & 2.01$\times 10^{36}$       & -              & -                         & $10.4\times13.3$   & -           & 37       &      -           & E(T)    \\
002.0$-$06.2 & M 2$-$33       & 18 15 06.52  & $-$30 15 33.05 & 0.27   & 1.22$\times 10^{32}$       & -              & O5f(H)                    & $15.1\times14.1$   & -           & 5.8      &       -          & E(T)    \\
002.1$-$08.3 & PPA1824$-$3107 & 18 24 21.74 & $-$31 07 09.51   & 0.25   & -              & -              & -                         & $8.7\times10.0$    & -           & -        & $7\times9$             & E(L)    \\
002.2$-$02.7 & M 2$-$23       & 18 01 42.72  & $-$28 25 45.05 & 0.88   & 7.36$\times 10^{32}$       & -              & Of                        & $11.1\times16.7$   & -           & 8.5      &      -           & E    \\
002.2$-$09.4 & Cn 1$-$5       & 18 29 11.57 & $-$31 29 59.24 & 0.23   & 3.45$\times 10^{31}$       & -              & [WO4]pec      & $18.2\times18.5$   & -           & 7        &     -            & R    \\
002.3$-$05.9 & PPA1814$-$2951 & 18 14 26.73 & $-$29 51 45.54 & 0.32   & -              & -              & -                         & $9.7\times13.8$    & -           & -        & $12\times14$           & E    \\
004.0$-$11.1 & M 3$-$29       & 18 39 25.77 & $-$30 40 36.59 & 0.15   & 1.00$\times 10^{31}$       & -              & -                         & $23.0\times20.2$   & -           & 8.2      &     -            & E    \\
004.9$-$08.6 & PPA1831$-$2849 & 18 31 11.20 & $-$28 49 13.73 & 0.32   & 3.52$\times 10^{32}$       & -              & -                         & $11.6\times11.9$   & -           & -        & $5\times8$             & R(L)    \\
005.1$-$08.9 & Hf 2$-$2       & 18 32 31.04 & $-$28 43 19.23 & 0.33   & 1.24$\times 10^{31}$       & -              & -                         & $18.0\times19.9$   & -           & 18.6     &     -            & E    \\
005.2$-$18.6 & StWr 2$-$21    & 19 14 23.37 & $-$32 34 17.63 & 0.1    & 2.53$\times 10^{31}$       & 5.96$\times 10^{32}$       & -                         & $11.2\times12.1$   & $8.5\times9.2$     & 5        & $2.7\times2.7$         & E    \\
005.8$-$09.2 & PPA1835$-$2811 & 18 35 20.99 & $-$28 11 52.18 & 0.26   & -              & -              & -                         & $13.4\times11.9$   & -           & -        & $11\times11$           & E(L)    \\
006.1$+$08.3 & M 1$-$20       & 17 28 57.64 & $-$19 15 54.06 & 0.84   & 1.44$\times 10^{32}$       & 2.05$\times 10^{34}$       & -                         & $8.0\times11.6$    & $8.7\times5.4$     & 7        & $2.5\times2.3$         & E    \\
006.8$-$19.8 & Wray 16$-$423  & 19 22 10.61 & $-$31 30 39.30 & 0.11   & 1.45$\times 10^{31}$       & 1.09$\times 10^{33}$       & [WC4-6]/wels  & $17.7\times17.0$   & $15.1\times13.6$   & -        & $1.4\times1.4$         & R    \\
007.9$+$10.1 & MaC 1$-$4      & 17 26 38.01 & $-$16 48 29.14 & 0.35   & -              & -              & -                         & $10.5\times11.9$   & 8           & -        &       -          & E    \\
008.2$+$06.8 & He 2$-$260     & 17 38 57.22 & $-$18 17 35.90 & 0.57   & 5.07$\times 10^{32}$       & -              & O                         & $23.1\times17.4$   & -           & 10       &      -           & E    \\
008.3$+$09.6 & PHR1729$-$1647 & 17 29 13.30 & $-$16 47 42.74 & 0.44   & -              & -              & -                         & $10.7\times16.2$   & -           & -        & $30\times14$           & E(T)    \\
011.0$+$05.8 & NGC 6439     & 17 48 19.92 & $-$16 28 44.20 & 0.67   & 1.75$\times 10^{32}$       & -              & -                         & $15.5\times12.4$   & -           & 5        &      -           & E    \\
012.5$-$09.8 & M 1$-$62       & 18 50 26.01 & $-$22 34 22.45 & 0.39   & 4.45$\times 10^{31}$       & 9.35$\times 10^{32}$       & wels                      & $12.9\times15.8$   & $14.8\times12.8$   & 3.6      & $4.8\times4.6$         & E    \\
013.3$+$32.7 & Sn 1         & 16 21 04.47  & 00 16 09.92   & 0.11   & 5.21$\times 10^{31}$       & 3.35$\times 10^{33}$       & -                         & $28.0\times30.1$   & $24.2\times29.8$   & 6        & $5.9\times5.0$         & E    \\
013.7$-$10.6 & Y$-$C 2$-$32     & 18 55 30.72 & $-$21 49 40.09 & 0.26   & -              & -              & wels                      & $18.2\times23.3$   & -           & 15       &        -         & E    \\
014.8$-$08.4 & PHR1849$-$1952 & 18 49 24.28 & $-$19 52 12.83 & 0.26   & -              & -              & -                         & $23.0\times24.0$   & -           & -        & $19\times17$           & R(T)    \\
014.8$-$25.6 & HDW 12       & 19 58 13.19 & $-$26 28 15.81 & 0.15   & -              & -              & -                         & $14.4\times15.8$   & $8.2\times10.5$    & 46       &        -         & E    \\
019.4$-$05.3 & M 1$-$61       & 18 45 55.16 & $-$14 27 37.56 & 0.7    & 4.60$\times 10^{31}$       & -              & wels                      & $9.7\times15.3$    & -           & -        &       -          & E    \\
027.6$-$09.6 & IC 4846      & 19 16 28.25 & $-$09 02 36.33   & 0.314  & 2.68$\times 10^{31}$       & 2.33$\times 10^{33}$       & Of                        & $24.8\times21.8$   & $19.0\times17.5$   & 2        & $3\times3$             & E    \\
028.0$+$10.2 & WeSb 3       & 18 06 00.72   & 00 22 39.13   & 0.44   & 5.78$\times 10^{31}$       & 5.77$\times 10^{32}$       & -                         & $14.4\times16.8$   & $16.8\times14.3$   & 36       & $50.7\times43.1$       & E    \\
032.1$+$07.0 & PC 19        & 18 24 44.50 & 02 29 28.35   & 0.76   & 1.90$\times 10^{32}$       & 2.31$\times 10^{33}$       & -                         & $8.5\times8.5$     & $8.1\times6.8$     & 14       & $3\times3$             & R    \\
032.9$+$07.8 & K 3$-$1        & 18 23 21.69 & 03 36 28.71   & 0.51   & 9.94$\times 10^{30}$       & 4.19$\times 10^{32}$       & -                         & $12.1\times11.7$   & $7.0\times6.5$     & 8        &      -           & R    \\
037.7$-$05.9 & PHR1921$+$0136 & 19 21 20.83 & 01 36 14.00   & 0.63   & -              & -              & -                         & $9.7\times7.3$     & -           & -        & $16\times16$           & E(T)    \\
037.8$-$06.3 & NGC 6790     & 19 22 56.93 & 01 30 46.06   & 0.49   & 9.75$\times 10^{30}$       & -              & WN?                       & $20.7\times16.8$   & -           & 7        &     -            & E    \\
038.4$-$03.3 & K 4$-$19       & 19 13 22.64 & 03 25 00.67    & 1.2    & -              & -              & -                         & $10.0\times11.6$   & -           & -        &      -           & E    \\
044.3$+$10.4 & We 3$-$1       & 18 34 02.61  & 14 49 20.51  & 0.35   & 1.54$\times 10^{30}$       & 5.77$\times 10^{31}$       & -                         & $14.8\times18.2$   & $14.6\times12.9$   & 135      & $175\times160$         & E    \\
054.4$-$02.5 & M 1$-$72       & 19 41 33.96 & 17 45 17.36  & 0.82   & 6.74$\times 10^{31}$       & -              & -                         & $6.5\times9.5$     & -           & 10       &       -          & E    \\
059.7$-$18.7 & A 72         & 20 50 02.15  & 13 33 28.93  & 0.099  & 5.73$\times 10^{30}$       & 6.71$\times 10^{31}$       & -                         & $23.0\times21.2$   & $20.9\times15.6$   & 127      & $154\times118$         & E    \\
061.9$+$41.3 & DdDm 1       & 16 40 18.20 & 38 42 20.56  & 0.012  & 5.29$\times 10^{31}$       & 3.40$\times 10^{33}$       & -                         & $27.6\times27.9$   & $20.2\times24.3$   & 0.6      & $1.4\times1.4$         & R    \\
066.7$-$28.2 & NGC 7094     & 21 36 52.93 & 12 47 20.97  & 0.136  & 7.96$\times 10^{29}$       & 5.63$\times 10^{32}$       & hybrid                    & $96.6\times105.1$  & $40.8\times44.7$   & 94.8     & $102.5\times99$        & E    \\
072.7$-$17.1 & A 74         & 21 16 52.31 & 24 08 51.49   & 0.14   & 2.71$\times 10^{28}$       & 8.02$\times 10^{29}$       & DAO                       & $20.4\times17.8$   & $15.8\times16.2$   & 830.4    & $828\times776$         & B    \\
081.2$-$14.9 & A 78         & 21 35 29.48 & 31 41 46.74  & 0.164  & 3.40$\times 10^{30}$       & 6.27$\times 10^{32}$       & [WC]-PG1159   & $70.2\times75.2$   & $59.9\times62.1$   & 107      & $128\times108$         & E    \\
082.5$+$11.3 & NGC 6833     & 19 49 46.42 & 48 57 39.08  & 0.14   & 4.00$\times 10^{32}$       & -              & Of                        & $30.3\times23.6$   & -           & 2        &                 & E    \\
083.5$+$12.7 & NGC 6826     & 19 44 48.35 & 50 31 30.52  & 0.11   & 3.02$\times 10^{31}$       & -              & O3f(H)                    & $94.9\times105.8$  & -           & 25       &                 & E    \\
084.9$-$03.4 & NGC 7027     & 21 07 01.55   & 42 14 10.41  & 1.226  & 2.63$\times 10^{33}$       & 9.66$\times 10^{34}$       & -                         & $32.6\times30.3$   & $29.8\times29.9$   & 14       & $15.6\times12$         & E    \\
089.8$-$05.1 & IC 5117      & 21 32 30.99 & 44 35 47.37  & 0.5    & 5.00$\times 10^{29}$       & -              & [WR]               & $9.7\times14.8$    & -           & 1.2      &                 & B    \\
096.4$+$29.9 & NGC 6543     & 17 58 33.66 & 66 37 58.15  & 0.05   & 3.44$\times 10^{32}$       & 7.01$\times 10^{32}$       & -                         & $115.3\times116.0$ & $101.7\times102.0$ & 19.5     & $26.5\times23.5$       & R    \\
100.0$-$08.7 & Me 2$-$2       & 22 31 43.72 & 47 48 04.84   & 0.18   & 6.25$\times 10^{31}$       & -              & Of                        & $29.9\times25.7$   & -           & 5        &                 & B    \\
100.6$-$05.4 & IC 5217      & 22 23 55.77 & 50 58 00.59   & 0.29   & 7.47$\times 10^{31}$       & -              & [WC8-9]?;wels & $35.7\times26.2$   & -           & 6.6      &                 & E    \\
106.5$-$17.6 & NGC 7662     & 23 25 53.87 & 42 32 05.54   & 0.14   & 2.75$\times 10^{31}$       & -              & UV em lines               & $52.2\times55.4$   & -           & 17       &                 & R    \\
107.6$-$13.3 & Vy 2$-$3      & 23 22 58.00 & 46 53 58.25  & 0.161  & 4.19$\times 10^{31}$       & 2.42$\times 10^{33}$       & -                         & $30.4\times32.1$   & $24.8\times30.6$   & 4.2      & $4.6\times4.6$         & R    \\
107.7$+$07.8 & IsWe 2       & 22 13 22.59 & 65 53 55.31  & 0.89   & 2.79$\times 10^{30}$       & -              & DA                        & $11.2\times9.7$    & -           & 900      &                 & E    \\
111.0$+$11.6 & DeHt 5       & 22 19 33.47 & 70 56 03.16   & 0.68   & 4.54$\times 10^{30}$       & -              & DA                        & $30.1\times25.3$   & -           & 528      &                 & E    \\
111.8$-$02.8 & Hb 12        & 23 26 14.92 & 58 10 54.30  & 2.15   & 1.47$\times 10^{36}$       & -              & B[e]?;WN7     & $16.8\times22.4$   & -           & 1        &                 & E    \\
119.6$-$06.1 & Hu 1$-$1       & 00 28 15.55  & 55 57 54.85  & 0.33   & 2.76$\times 10^{31}$       & -              & DA                        & $18.4\times18.4$   & -           & 5        &                 & R    \\
120.0$+$09.8 & NGC 40       & 00 13 01.16   & 72 31 19.76  & 0.39   & 3.19$\times 10^{30}$       & 2.91$\times 10^{33}$       & DAO                       & $68.2\times73.6$   & 18          & 48       & $56\times34$           & E    \\
130.2$+$01.3 & IC 1747      & 01 57 35.61  & 63 19 17.20  & 1.15   & 4.41$\times 10^{33}$       & -              & [WO]                       & $14.3\times17.5$   & -           & 13       &                 & E    \\
130.3$-$11.7 & M 1$-$1        & 01 37 19.42  & 50 28 11.08  & 0.2    & 2.70$\times 10^{30}$       & -              & -                         & $19.9\times15.8$   & -           & 6        &                 & E    \\
\hline
\end{tabular}
\begin{tablenotes}
\item The values in parenthesis are for sources that are common with MASH catalogue \citep{Parker06}. The objects are classified as true PN (T), probably PN (P) and likely PN (L), as per the catalogue.
\end{tablenotes}
}
\end{table}
\end{landscape}

%% file: table4.tex

\begin{landscape}
\begin{table}

\scriptsize{
\begin{tabular}{ccccccccccccc}
{Table~\ref{tab1} } & &&&&&&\\
\hline
PNG & Name & RA (J2000) & Dec (J2000) & E(B-V) & L\_NUV & L\_FUV & ST of CSPN & NUV\_size & FUV\_size & V\_size & H$\alpha$\_size & type \\
 &  & (hh mm ss) & (hh mm ss) & & (erg s$^{-1}$) & (erg s$^{-1}$) &  & (arcsec) & (arcsec) & (arcsec) & (arcsec) &  \\
\hline

138.8$+$02.8 & IC 289       & 03 10 19.44  & 61 19 03.36   & 1.24   & 1.29$\times 10^{34}$       & -              & OH                                  & $9.4\times10.5$   & -         & 35       &                 & E    \\
146.7$+$07.6 & M 4$-$18       & 04 25 50.77  & 60 07 13.50   & 0.61   & 3.94$\times 10^{32}$       & -              & [WC11]                                & $17.5\times20.1$  & -         & 4        &                 & E    \\
153.7$+$22.8 & A 16         & 06 43 55.37  & 61 47 22.67  & 0.15   & 4.37$\times 10^{29}$       & 6.06$\times 10^{30}$       & -                                   & $13.1\times13.4$  & $13.3\times13.1$ & 105      & $148\times140$         & R    \\
159.0$-$15.1 & IC 351       & 03 47 32.96  & 35 02 49.61   & 0.24   & 5.68$\times 10^{30}$       & -              & wels                                & $19.9\times22.3$  & -         & 7        &                 & E    \\
161.2$-$14.8 & IC 2003      & 03 56 21.93  & 33 52 30.05  & 0.24   & 1.90$\times 10^{31}$       & -              & [WC3]                                 & $19.2\times31.5$  & -         & 8.6      &                 & E    \\
165.5$-$06.5 & K 3$-$67       & 04 39 47.90  & 36 45 43.37  & 0.81   & 3.15$\times 10^{31}$       & 2.72$\times 10^{32}$       & O(C);CIV+He em                      & $9.7\times9.7$    & $8.7\times8.8$   & -        &                 & R    \\
165.5$-$15.2 & NGC 1514     & 04 09 17.11   & 30 46 33.35  & 0.68   & 6.34$\times 10^{31}$       & 1.06$\times 10^{33}$       & sdO+A0 III                          & $39.6\times37.6$  & $30.1\times24.5$ & 132      & $188\times182$         & R    \\
166.1$+$10.4 & IC 2149      & 05 56 24.04  & 46 06 19.38   & 0.23   & 3.17$\times 10^{32}$       & -              & O4f                                 & $53.6\times45.7$  & -         & 8.5      &                 & E    \\
167.4$-$09.1 & K 3$-$66       & 04 36 37.27  & 33 39 30.33  & 0.78   & 4.27$\times 10^{31}$       & 1.33$\times 10^{33}$       & cont.                               & $14.0\times10.2$  & $8.7\times8.0$   & -        & $2.1\times2.1$         & E    \\
170.3$+$15.8 & NGC 2242     & 06 34 07.30   & 44 46 37.93  & 0.132  & 3.17$\times 10^{30}$       & 6.59$\times 10^{32}$       & O(H)                                & $31.3\times32.6$  & $28.6\times33.8$ & 22       & $20\times20$           & R    \\
190.3$-$17.7 & J 320        & 05 05 34.30   & 10 42 22.26  & 0.208  & 2.00$\times 10^{31}$       & 7.06$\times 10^{32}$       & wels                                & $24.0\times25.3$  & $22.1\times20.4$ & 6.4      & $9.4\times6.3$         & R    \\
208.5$+$33.2 & A 30         & 08 46 53.48  & 17 52 47.61  & 0.023  & 1.63$\times 10^{30}$       & -              & -                                   & $51.7\times65.3$  & -         & 127      &                 & E    \\
211.4$+$18.4 & HDW 7        & 07 55 11.29  & 09 33 09.62    & 0.02   & -              & -              & hgO(H)                              & $17.0\times18.5$  & $17.7\times18.5$ & 94       & $105\times105$         & E    \\
222.1$-$03.9 & PFP1         & 07 22 17.81  & $-$06 21 45.46  & 0.26   & 6.87$\times 10^{29}$       & -              & pre-WD?                             & $25.0\times26.87$ & -         & -        & $1150\times1100$       & E(T)    \\
226.7$+$05.6 & M 1$-$16       & 07 37 18.99  & $-$09 38 47.92  & 0.22   & -              & -              & -                                   & $17.2\times18.0$  & -         & 3        &                 & E    \\
228.2$-$22.1 & DeHt 1       & 05 55 06.58   & $-$22 54 02.19  & 0.041  & 2.55$\times 10^{30}$       & -              & KV                                  & $20.7\times19.4$  & $16.8\times16.3$ & 132      & $142\times142$         & R    \\
231.8$+$04.1 & NGC 2438     & 07 41 50.32  & $-$14 44 06.65  & 0.37   & 1.86$\times 10^{30}$       & -              & O(H)                                & $62.6\times66.2$  & -         & 64       &                 & B    \\
234.8$+$02.4 & NGC 2440     & 07 41 55.27  & $-$18 12 32.64 & 0.61   & 8.31$\times 10^{31}$       & -              & -                                   & $70.1\times46.8$  & -         & 16       &                 & E    \\
235.7$+$07.1 & PHR0800$-$1635 & 08 00 59.09   & $-$16 35 37.32 & 0.11   & 3.50$\times 10^{29}$       & -              & -                                   & $11.7\times9.0$   & -         & -        & $157\times150$         & E(T)    \\
239.6$+$13.9 & NGC 2610     & 08 33 23.44  & $-$16 08 57.42  & 0.055  & 3.18$\times 10^{30}$       & 4.52$\times 10^{32}$       & WD?                                 & $58.0\times59.0$  & $64.4\times58.3$ & 38       & $49.7\times47.6$       & B    \\
243.3$-$01.0 & NGC 2452     & 07 47 26.23  & $-$27 20 04.29  & 1.15   & 2.58$\times 10^{33}$       & -              & [WO 1]                   & $22.8\times37.8$  & -         & 19       &                 & E    \\
243.8$-$37.1 & PRTM 1       & 05 03 01.64    & $-$39 45 43.83 & 0.026  & 1.66$\times 10^{31}$       & 1.87$\times 10^{33}$       & O(H)                                & $26.5\times29.8$  & $28.4\times25.0$ & 23       & $21.3\times20.5$       & E    \\
249.0$+$06.9 & SaSt 1$-$1     & 08 31 42.99  & $-$27 45 31.99 & 0.14   & -              & -              & G2III                               & $29.8\times15.1$  & -         & -        &                 & B    \\
249.8$+$07.1 & PHR0834$-$2819 & 08 34 18.13  & $-$28 19 03.26  & 0.13   & 3.09$\times 10^{29}$       & -              & -                                   & $13.6\times10.7$  & -         & -        & $160\times142$         & E(T)    \\
255.8$+$10.9 & FP0905$-$3033  & 09 05 05.37    & $-$30 33 13.19 & 0.222  & 1.13$\times 10^{29}$       & 1.04$\times 10^{31}$       & -                                   & $19.2\times20.7$  & $16.3\times16.8$ & -        & $882\times660$         & E(L)    \\
270.1$+$24.8 & K 1$-$28       & 10 34 30.64 & $-$29 11 14.16 & 0.052  & 7.82$\times 10^{30}$       & 1.80$\times 10^{32}$       & [WO 4]-[WC 4] & 23.5$\times$20.9  & $27.7\times21.4$ & 54       & $54\times47$           & E    \\
286.8$-$29.5 & K 1$-$27       & 05 57 01.94   & $-$75 40 22.46 & 0.079  & 5.93$\times 10^{30}$       & 5.75$\times 10^{32}$       & O(He)                               & $40.6\times35.5$  & $25.8\times25.7$ & 46       & $61\times47$           & E    \\
291.3$+$08.4 & PHR1134$-$5243 & 11 34 38.59 & $-$52 43 31.92 & 0.28   & 6.77$\times 10^{31}$       & -              & -                                   & $43.2\times34.34$ & -         & -        & $39\times39$           & E(T)    \\
296.4$-$06.9 & He 2$-$71      & 11 39 11.24 & $-$68 52 09.73  & 0.4    & 1.28$\times 10^{32}$       & -              & O(H)                                & $21.4\times25.5$  & -         & 5        &                 & E    \\
308.2$+$07.7 & MeWe 1$-$3     & 13 28 04.80  & $-$54 41 58.31 & 0.42   & 3.08$\times 10^{32}$       & -              & -                                   & $15.6\times14.1$  & -         & 18       &                 & E    \\
309.0$-$04.2 & He 2$-$99      & 13 52 30.65 & $-$66 23 27.26 & 0.5    & 5.93$\times 10^{32}$       & -              & -                                   & $19.6\times28.7$  & -         & 17       &                 & B    \\
310.3$+$24.7 & Lo 8         & 13 25 37.23 & $-$37 36 15.28 & 0.07   & 2.79$\times 10^{31}$       & -              & DAO                                 & $57.8\times67.3$  & -         & 115      &                 & E    \\
311.7$+$07.3 & PHR1351$-$5429 & 13 51 46.55 & $-$54 29 42.45 & 0.45   & 5.63$\times 10^{31}$       & -              & [WC 9]                   & $9.2\times5.7$    & -         & -        & $36\times35$           & E(T)    \\
316.3$+$08.8 & PHR1418$-$5144 & 14 18 25.86 & $-$51 44 38.81 & 0.38   & 1.64$\times 10^{30}$       & -              & -                                   & $19.4\times13.3$  & -         & -        & $404\times375$         & E(T)    \\
320.6$-$04.8 & PHR1532$-$6203 & 15 32 59.39 & $-$62 03 03.39   & 0.52   & 2.88$\times 10^{32}$       & -              & WD;sdO                               & $44.2\times41.2$  & -         & -        & $15.5\times16$         & E(T)    \\
325.8$-$12.8 & He 2$-$182     & 16 54 35.36 & $-$64 14 29.95 & 0.16   & 2.30$\times 10^{32}$       & -              & -                                   & $18.0\times29.4$  & -         & 3        &                 & E    \\
327.7$-$05.4 & KoRe 1       & 16 19 15.55 & $-$57 58 24.57 & 0.34   & 1.64$\times 10^{34}$       & -              & -                                   & $37.8\times43.2$  & -         & 14.2     &                 & B    \\
330.6$-$03.6 & He 2$-$159     & 16 24 21.32 & $-$54 36 03.21  & 0.59   & 8.40$\times 10^{31}$       & -              & O(H)                                & $16.2\times13.9$  & -         & 10       &                 & E    \\
334.3$-$09.3 & IC 4642      & 17 11 45.19 & $-$55 24 02.42  & 0.19   & 9.72$\times 10^{31}$       & -              & -                                   & $21.4\times20.2$  & —         & 16.5     &                 & R    \\
336.8$-$07.2 & K 2$-$17       & 17 09 35.95  & $-$52 13 03.76  & 0.37   & 1.34$\times 10^{31}$       & -              & abs. lines                          & $43.4\times44.9$  & —         & 38       &                 & R    \\
340.4$-$14.1 & Sa 1$-$6       & 18 00 59.43  & $-$52 44 20.17 & 0.132  & -              & -              & Of                                  & $26.9\times27.9$  & $25.0\times30.8$ & 11       &                 & R    \\
341.6$+$13.7 & NGC 6026     & 16 01 20.91  & $-$34 32 37.87 & 0.35   & 1.00$\times 10^{31}$       & 2.63$\times 10^{33}$       & -                                   & $10.9\times13.4$  & 15        & 40       & $53\times45.5$         & E    \\
342.1$+$27.5 & Me 2$-$1       & 15 22 19.20 & $-$23 37 32.07 & 0.14   & 3.84$\times 10^{32}$       & -              & -                                   & $6.1\times9.0$    & -         & 6        &                 & E    \\
344.4$-$06.1 & Wray 16$-$278  & 17 30 03.87  & $-$45 22 51.46 & 0.43   & -              & -              & -                                   & $6.1\times12.2$   & -         & -        &                 & E    \\
346.9$-$08.9 & PHR1750$-$4438 & 17 50 45.70 & $-$44 38 21.26 & 0.29   & -              & -              & -                                   & $11.6\times17.8$  & -         & -        & $22\times20$           & E(T)    \\
351.3$+$07.6 & H 1$-$4        & 16 53 37.07 & $-$31 40 32.82 & 0.4143 & -              & -              & -                                   & $38.3\times22.8$  & -         & -        &                 & E    \\
351.9$+$09.0 & PC 13        & 16 50 17.08 & $-$30 19 54.21 & 0.37   & 5.17$\times 10^{31}$       & -              & -                                   & $15.0\times15.3$  & -         & 7        &                 & R    \\
353.3$+$06.3 & M 2$-$6        & 17 04 18.32  & $-$30 53 29.34 & 0.72   & 1.38$\times 10^{30}$       & -              & -                                   & $14.8\times17.0$  & -         & 8        &                 & E    \\
353.7$-$12.8 & Wray 16$-$411  & 18 26 41.60  & $-$40 29 53.83 & 0.09   & 4.02$\times 10^{31}$       & -              & -                                   & $11.6\times14.8$  & -         & 30       &                 & B    \\
355.7$-$03.5 & H 1$-$35       & 17 49 13.93 & $-$34 22 53.19 & 0.85   & 2.64$\times 10^{32}$       & -              & -                                   & $9.0\times11.0$   & -         & 2        &                 & B    \\
356.5$-$02.3 & M 1$-$27       & 17 46 45.50 & $-$33 08 34.61  & 1.92   & 1.88$\times 10^{30}$       & -              & -                                   & $10.2\times10.4$  & -         & 5.3      &                 & E    \\
356.8$-$03.6 & PHR1752$-$3330 & 17 52 29.23 & $-$33 30 05.37  & 0.99   & -              & -              & wels                                & $24.0\times27.9$  & -         & -        & $26\times22$           & R(P)    \\
358.3$-$21.6 & IC 1297      & 19 17 23.37 & $-$39 36 45.79 & 0.118  & 9.91$\times 10^{33}$       & 2.73$\times 10^{35}$       & [WO 3]                   & $11.7\times14.8$  & 19        & 7        & $10.8\times9.8$        & E    \\
358.5$-$04.2 & H 1$-$46       & 17 59 02.52  & $-$32 21 43.84 & 0.79   & 2.66$\times 10^{34}$       & 3.52$\times 10^{34}$       & -                                   & 24         & -         & -        &                 & R    \\
358.9$-$00.7 & M 1$-$26       & 17 45 57.64 & $-$30 11 59.98 & 9.06   & -       & -              & [WC 11]?                 & 15         & -         & 4.2      &                 & R\\
\hline
\end{tabular}
\begin{tablenotes}
\item The values in parenthesis are for sources that are common with MASH catalogue \citep{Parker06}. The objects are classified as true PN (T), probably PN (P) and likely PN (L), as per the catalogue.
\end{tablenotes}
}
\end{table}

\end{landscape}